\documentclass[english,12pt,aps,prd,eqsecnum,nofootinbib]{revtex4}
\usepackage{amssymb,amsmath,amsthm,graphicx,amscd}
\usepackage[dvipsnames]{xcolor}
\usepackage{esint,enumerate,comment,ulem}
\usepackage{babel}
\usepackage[hidelinks]{hyperref}



\begin{document}

\title{Ground State Excitation of an Atom Strongly Coupled \\ to a Free Quantum Field}

\author{Jen-Tsung Hsiang$^{1}$}%
\thanks{cosmology@gmail.com}%
\author{Bei-Lok Hu$^{2}$}%
\thanks{blhu@umd.edu}

\affiliation{$^{1}$ Center for High Energy and High Field Physics, National Central University, Chungli 32001, Taiwan}

\affiliation{$^{2}$ Joint Quantum Institute and Maryland Center for Fundamental Physics,\\ University of Maryland, College Park, Maryland 20742, USA}

\begin{abstract}
This paper presents a nonperturbative treatment of strong-coupling induced effects in atom-field systems which cannot be seen in traditional perturbative treatments invoking  compromising assumptions such as the Born-Markov, rotating wave or Fermi Golden rule. We consider an atom whose internal degrees of freedom are modeled by a harmonic oscillator  which is bilinearly coupled to a scalar quantum field, representing one of the two polarizations of an electromagnetic field. Because the whole system is Gaussian we can solve this problem exactly. Using the open quantum system conceptual framework and the influence functional formalism we derive the dynamics of the reduced density matrix for the atom which enables the calculation of atomic transition probability and other relevant physical quantities. Finding an exact solution to this problem has the distinct advantage of enabling one to capture fully the strong coupling regime  and discover interesting effects such as spontaneous ground state excitation~\cite{PPP95} which is unfathomable in perturbative treatments. The conventional description of atomic-optical activities is predicated on the assumption that the state of the total atom-field system is a product state of the atomic excitations and the photon number states, an assumption which is valid only for vanishingly weak coupling. The correct energy eigenfunctions to use should be that of the Hamiltonian of the combined atom-field system. Other features associated with finite to strong coupling effects such as resonance peak broadening and transition from a gapped to a gapless spectrum can all be understood from this perspective. Finally, to put the issues in a proper perspective we  take the perturbative limit of the exact results and compare them with those from conventional time-dependent perturbation theory (TDPT). This enables one to pin-point where the deficiencies of TDPT lie as one removes the ultra-weak coupling assumption.     
\end{abstract}

\maketitle
\baselineskip=18pt
\allowdisplaybreaks

\section{Introduction}

Interaction between atoms and light \cite{AFI,SA10} forms one important  cornerstone of modern physics, serving as the foundation of contemporary atomic-optical physics with wide ranging applications, while illuminating fundamental theoretical issues in the structure of matter as embodied fully  in quantum electrodynamics (QED). Cavity (C) QED \cite{CQED} has proven to be very successful in gaining precision control of few single atoms interacting with a finite number of field modes. Traditional operating parameters of CQED fall largely in the weak coupling range  (see, the table in \cite{DR18}). The attempt toward achieving strong coupling between atom and field in a cavity started about two decades ago (e.g., \cite{Ki98,EK01,MilKim05,DG09}) while ultrastrong coupling was realized soon after in circuit (c) QED \cite{BH04,WS04,SG08,ND06,AA08,FG17,YF17}. 

Instead of cavities, strong coupling between an atom and a quantum field in free space, where an infinite number of modes can in principle partake of, invokes many new challenges but also offers new opportunities~\cite{KM19,AtomPhoton,Tey08,AZ10,Tey09,LS13}  in atomic-optical physics, from quantum communication (e.g., with emitter of radiation serving as the interface between stationary qubits in matter and flying qubits like photon \cite{SAL09}) to  physio-biology (e.g., excitation transport in photosynthetic complexes \cite{Levi15}).

{Generically the effective atom-field coupling $\mathsf{g}$ takes the form
\begin{equation}
	\mathsf{g}\sim d\sqrt{\frac{N\omega}{V_{\textsc{q}}}}\,,
\end{equation}
where $d$ is the atomic transition dipole moment, $V_{\textsc{q}}$ is the quantization volume, $\omega$ the light (angular) frequency, and $N$ the number of the atoms involved~\cite{DR18}. Experimental realization of strong atom-field interaction in free space focuses on optimizing these parameters to achieve a large effective coupling strength. Among the most common approaches, one may 1) focus a laser beam on the location of the atom, 2) increase the effective dipole moment of the atom, or 3) form a large cluster of atoms in collective modes. Though conceptually intuitive, experimental implementations still pose great challenges.

\paragraph{Experimental aspects.} Experimentally the challenge begins with the difficulty in focusing light on a single atom. As explained in \cite{EK01} 
since light beams are transversely polarized, only part of the  {tightly focused} light entering the interaction region will carry the polarization that the atom is sensitive to.  
In an experiment the design should be such that the radiation incident onto the atom resembles {the emitted} dipole wave, 
since this is  the mode of the electromagnetic field that maximizes the electric field  
at the location of the atom~\cite{SL13,vN04}.   A comparison between the coupling scheme of cavity QED and the free space scenario is provided in \cite{LeuSon13}. 

Spontaneous emission from an atom in free space is isotropic. One can change the properties of the electromagnetic vacuum surrounding the atom to create directionality. However, mode matching between the input field and the dipolar emission pattern of the quantum emitter in free space is not easy and can limit the achievable coupling strength. 
Waveguide QED systems seek to overcome this limitation by transversely confining the propagating EM mode coupled to one or more emitters. 

Another approach is by {\it superradiance of an atom assembly}.  The collective interaction of all atoms with the available spatial vacuum modes sets the emission directionality. One can even design `superatoms' to scale up the coupling strength.  (See \cite{Mi14,PM15} for a review of this topic.) Recently  Paris et al \cite{PM17} reported on the realization of coherent coupling  between a propagating few-photon optical field and a single Rydberg superatom in free space. By exploiting the Rydberg blockade effect in an atomic ensemble 
which allows only a single excitation shared among all $N$ constituents, these authors turn about 104 individual ultracold atoms into a single effective two-level quantum system.  The collective nature of this excitation enhances the coupling of the light field to the superatom by a factor of $\sqrt{N}$ compared to the single-atom coupling strength and guarantees an enhanced directed emission in the forward direction. 

Another assembly where strong coupling has been achieved \cite{Huppert16,VT16} is in a very dense two-dimensional electron gas confined in semiconductor quantum wells. The collective mode, known as multi-subband plasmon, has a superradiant nature. 
Radiative lifetimes as short as $10\,\mathrm{fs}$ have been reported, thus much shorter than any nonradiative scattering process in the structure. As the radiative broadening is larger than the nonradiative one, the collective excitation can be considered as strongly coupled with free-space radiation.

While experimental advances in free space atom-field interactions hold huge potential for new applications  we are drawn to this problem because of new challenges in the underlying theoretical issues, namely, \textit{strong coupling, self-consistent back-action and non-Markovian effects}. 

\paragraph{Theoretical Issues} We mention three group of issues which require innovative treatment and analysis. A) At strong coupling many assumptions  made in conventional quantum optics such as the rotating-wave (RW)  and Born-Markov (BM) approximations are no longer valid~\cite{CH08}. Contributions from the anti-resonant/counter-rotating terms from the $\mathbf{p}\cdot\mathbf{A}$ expression~\cite{Mi94,CP95} and the diamagnetic $\mathbf{A}^{2}$ term ~\cite{MT16,VG14,GP15,VG15,CR10} in the atom-field Hamiltonian cannot be ignored. B) At strong coupling  the back-action of the field on the atom becomes strong, and  a self-consistent treatment of  the atom-field system is required.  C) Because  backaction brings in different time scales from the environment the evolution of the atomic system will have multi-time features and dependence on its past history, thus non-Markovian effects need be included in the consideration. 

In this paper we work with a harmonic atom strongly coupled to a quantum scalar field (which represents each of the two polarizations of the EM field) {in free space}, i.e. the atom's internal degrees of freedom are described by a harmonic oscillator, not a two level system\footnote{We should not forget that the oft-used ``two level atom'' is also an idealization. Two levels are selected out for a simple enough yet sufficiently accurate depiction of the specific physics of interest to the modeler, such as energy exchange with a cavity field mode in quantum optics or acting as a qubit in quantum information processing. A harmonic atom at very low temperature, when only the first excited state has some finite probability to be occupied, can act effectively like a two-level atom \cite{Shiokawa}.}.
The theoretical model for a two-level atom strongly interacting with one or several EM modes is described by the celebrated quantum Rabi model \cite{qRabi}, where exact solutions exist \cite{Braak11} for it and its generalizations (e.g., \cite{Eck17}). The special case of a single cavity mode under the rotating wave approximation is the famous Jaynes-Cummings model \cite{JC63}. Many papers on the quantum Rabi model have appeared in recent years, including a special issue \cite{Rabi80} and two topical reviews \cite{Xie17,Font-Diaz},  which we refer the reader to. With a harmonic atom coupled with arbitrary strength to a linear field  of a continuous spectrum, the combined system is Gaussian, whereby  exact solutions can be found. Exact solutions are highly desirable as they reveal fully the true physics which is likely marred or misled by whatever approximation schemes introduced. For this purpose, earlier  Karpov et al \cite{KPPP} sought exact solutions to the Friedrichs model with virtual transitions; Ciccarello et al \cite{CKP} presented an exactly solvable model of two three-dimensional harmonic oscillators interacting with a quantum electromagnetic field and obtained the far-zone Casimir-Polder potential. Passante et al~\cite{PRSTP} used the harmonic oscillator model to calculate the atom-surface Casimir-Polder interaction energy.

Below we describe several interesting features in our results related to strong coupling based on this harmonic atom-scalar field model. We shall then provide the details of our methodology and calculation.

\subsection{New Features of Strong Atom-Field Coupling}

\paragraph{Resonance Peak Broadening}

 An immediate effect of  strong atom-field coupling is the broadening of the resonance peak of the harmonic atom's response to   quantum fluctuations of the field. This effect shows up already in a classical driven oscillator where a larger damping constant of the oscillator will yield a wider resonance response to the diving force. As the damping constant which is proportional to the coupling constant increases, the resonance shape gradually becomes ill-defined, even unrecognizable. This implies that at sufficiently strong coupling and damping, the discrete energy-level distribution of the harmonic atom will  morph into a quasi-continuous distribution~\cite{HZ95}. This will bring up interesting physics which we shall return in the latter part of this paper.

\paragraph{Product state not an eigenstate of the total Hamiltonian}

In the context of atom-field interaction, we often describe the state of the total system in terms of the product states of the atomic excitations and the photon number states. This description offers vivid physical pictures of various quantum optical phenomena in terms transition of atomic levels by absorbing or emitting photons of the electromagnetic fields. In truth, such a product state is usually not an eigenstate of the total system, but it is close enough when the coupling between the atom and field is sufficiently weak such that the contribution from the interaction term in the total atom-field Hamiltonian is relatively small, compared to the free atom and the free field Hamiltonians. As the coupling strength increase, the deviation between the product state and the eigenstate of the total Hamiltonian becomes more and more discernible.  With finite strength interaction new features emerge.  Expressing  the combined state of an atom with (some modes of) an electromagnetic field in terms of a \textit{dressed atom}  is an old and useful concept~\cite{Mi94,CP95,FHMal}, as is the use of a \textit{polaron or polariton} basis to describe an electron  or atom interacting with a lattice (e.g.,~\cite{polaron,polariton}) or  a particle moving in a dielectric medium (e.g., \cite{Cao}).  Canonical quantization of the total system of an atom in a dielectric obtained by Fano quantization \cite{Fano} has long been studied (e.g.,~\cite{HutBar,Suttorp,Eberlein}. Bogoliubov-like transformations have been used for diagonalizing\footnote{A note on  instantaneous diagonalization of the Hamiltonian (e.g., \cite{GriMam}).  Beware of the pitfalls when applying this method to dynamically evolving systems~\cite{Fu79}  -- this was long known in quantum field theory in dynamical spacetimes~\cite{FuQFT} as applied to dynamical quantum processes such as cosmological particle creation or  dynamical Casimir effect.  The vacuum state defined at any instant of time is different from  the vacuum defined at another instant of time.  The corresponding Fock spaces of quantum field theories being inequivalent will give the wrong prediction of infinite particle production. The existence of a global time-like Killing vector is a necessary condition for a well-defined vacuum state~\cite{Ful73}. Given a well defined initial state and a final state if one is only interested in the amount of total particles produced at late times one can treat this problem by S-matrix (`in-out') techniques.  Otherwise one needs to use an adiabatic vacuum \cite{ParFul74,Hu74,FulParHu74} and follow the proper procedures, such as for the regularization of the quantum stress energy tensor, etc.} the interacting Hamiltonian of the system~\cite{KPPP}. These entities and methods are conceptually and formally valid in the strong coupling limit; the challenge is to identify qualitatively new effects without using any approximation.  


\paragraph{Spontaneous ground state excitation}

It is well known that if we prepare the atom-field system initially in one of the product states, then since such an initial state is not a total energy eigenstate, the combined system will not stay stationary but evolve  to a state different from the initial one. However, what is perhaps lesser known or talked about is that the product of the atomic ground state and the vacuum state of the field may not be the lowest energy state of the total system and that the ground-state energy of the reduced oscillator system~\cite{NB02} can fluctuate. The former feature has been discussed in the context of quantum entanglement in the oscillator-field system~\cite{JB04,JB05,BJ05} (see also \cite{AngPaz95,LinHu06,LinHu07}). While, in particular, the latter shows that, in general, the ground state energy of the reduced oscillator system can fluctuate. This ground-state energy fluctuation is vanishingly small when the oscillator-field coupling strength is diminutive, as is commonly believed. However, as the coupling strength increases, there will be a crossover to the case when the ground-state energy uncertainty becomes greater than the mean value. 

Historically Passante, Petrosky and Prigogine~\cite{PPP95} while studying a two level system interacting with  a radiation field suggested that a  virtual transition can first bring the system to a higher excited level, from which it subsequently decays by a resonant transition. They called this virtual transition a case of `indirect spectroscopy'. Because exact solutions are not available for two level system-field interactions, some approximation is bound to be necessary. Arguments similar to what we presented above have been invoked to explain  processes in atomic systems where  virtual photons {in the ground state of the combined system} are  converted to real photons~\cite{SR13,DL17} and the so-called ground-state electroluminescence effect~\cite{CL16}. 
Our calculation shows that even when the combined system is initially prepared in the product state of the atomic ground state and the quantum field vacuum, if we perform a projective measurement, with respect to the original free states,  of the final equilibrium state of the reduced atomic system, we will find a nonzero probability that the atom shows up in the excited states.  Our results  concretize this phenomenon of spontaneous ground state excitation as a real physical process in strongly coupled atom-field systems.

The paper is organized as follows: The next section describes the open system quantum dynamics methodology for the calculation of the transition probability using the framework of reduced density matrix, in the Feynman-Vernon influence functional (IF) \cite{CalHu,FeyVer,CalLeg,HPZ} and its close kin, the  `in-in', Schwinger-Keldysh or `closed-time-path'(CTP) \cite{CTP}  formalism. 
We emphasize that transition probability calculated by means of time-dependent perturbation theory or under any Markovian approximation such as the Fermi Golden rule or using the Lindblad, Redfield types of master equations or their associated Langevin or Fokker-Planck equations are categorically inadequate to capture the full fledged features of strong coupling effects. In Sec.~\ref{S:erbffe}, we introduce ``spontaneous excitation'', a seemingly paradoxical phenomenon when the atom-field coupling is sufficiently strong and use this as an example to highlight certain shortcomings in time-dependent perturbation theory. We then explain the physics behind the ``spontaneous excitation''. In Sec.~\ref{S:nriuthdf}, we discuss atomic transition from its first excited state, and offer a qualitative analysis of the late-time transition probability from the first excited level to the neighboring levels. There we can see the general features of the dependence of the transition probability on the atom-field coupling strength and the connection with the traditional weak-coupling transition.  A more quantitative investigation is given in Sec.~\ref{S:rtkbgfhgse}, where we discuss how different parameters in the theory enter in the nonequilibrium, transient dynamics of the transition probability at strong coupling. In the Appendix, we present in detail a comparison and the connection between the results obtained from the `in-in' formalism and  time-dependent perturbation theory.


\section{Open Quantum System Dynamics}

Our goal is to calculate the transition probability and spontaneous emission of an atom  strongly coupled to a quantum field. Traditional methods deployed under the ultra-weak coupling assumption such as time-dependent perturbation theory, or under any Markovian approximation such as the Fermi Golden rule, or the use of Lindblad or Redfield type of master equations are inadequate for the investigation of strong coupling effects. Instead we focus on the reduced density matrix in the Feynman-Vernon (influence functional) formalism with Schwinger-Keldysh (closed-time-path) techniques, as used earlier in \cite{BehHu10,BehHu11} for atom-field and atom-dielectric-field interactions, and in our forthcoming main series of papers on atom-field-medium interactions \cite{AFM02,AFM1}.  The influence functional method has also been used for tackling Casimir \cite{Lombardo}, dynamical Casimir effects and quantum friction \cite{Frias}.  We shall describe how  the `in-out' transition probability relevant for our purpose here can be phrased in the `in-in' language.

\subsection{Density Matrix}

The dynamics of a quantum system can be fully described by the density matrix operator $\hat{\varrho}$ because the quantum expectation value of any physical variable operator $\hat{\mathcal{O}}$ is given by
\begin{equation}
	\langle\hat{\mathcal{O}}\rangle=\operatorname{Tr}\Bigl\{\hat{\mathcal{O}}\,\hat{\varrho}\Bigr\}\,.
\end{equation}
The time evolution of the density matrix on the other hand is mapped by the superoperator $\hat{\mathcal{J}}$ from an initial time $t_{i}$ to a final time $t_{f}$
\begin{equation}\label{E:trbkjbfg}
	\hat{\varrho}(t_{f})=\hat{\mathcal{J}}(t_{f},t_{i})\circ\hat{\varrho}(t_{i})\,,
\end{equation}
which is realized by the unitary transformation of the density matrix
\begin{equation}
	\hat{\varrho}(t_{f})=\hat{U}(t_{f},t_{i})\cdot\hat{\varrho}(t_{i})\cdot \hat{U}^{\dagger}(t_{f},t_{i})\,,
\end{equation}
via the time evolution operator $\hat{U}(t_{f},t_{i})$, which takes the form
\begin{equation}\label{E:jbhrtr}
	\hat{U}(t_{f},t_{i})=\operatorname{T}_{+}\exp\biggl[-i\int_{t_{i}}^{t_{f}}\!dt\;\hat{H}(t)\biggr]\,,
\end{equation}
where $\hat{H}$ is the Hamiltonian operator of the quantum system, and $\operatorname{T}_{+}$ denotes time-ordering.

\subsection{Transition Probability}

The transition probability of the quantum system can also be formulated in terms of the density matrix. Since the transition probability from an initial state $\lvert\psi_{i}\rangle$ to a final state $\lvert\varphi_{f}\rangle$ is defined by
\begin{equation}
	P_{i\to f}=\Bigl|\langle\varphi_{f}\vert \hat{U}(t_{f}-t_{i})\vert\psi_{i}\rangle\Bigr|^{2}\,,
\end{equation}
it can be cast into
\begin{equation}\label{E:gksjgsx}
	P_{i\to f}=\operatorname{Tr}\Bigl\{\lvert\varphi_{f}\rangle\langle\varphi_{f}\rvert\,\hat{\varrho}_{\psi}(t_{f})\Bigr\}
\end{equation}
where the density matrix $\varrho_{\psi}$ at $t=t_{f}$, which has evolved from the initial state $\lvert\psi_{i}\rangle$, is given by
\begin{equation}
	\hat{\varrho}_{\psi}(t_{f})=U(t_{f},t_{i})\,\lvert\psi_{i}\rangle\langle\psi_{i}\rvert\,\hat{U}^{\dagger}(t_{f},t_{i})=\hat{U}(t_{f},t_{i})\,\hat{\varrho}_{\psi}(t_{i})\,\hat{U}^{\dagger}(t_{f},t_{i})\,,
\end{equation}
according to \eqref{E:trbkjbfg}. That is, the transition probability in fact is the expectation value of the projection operator $\lvert\varphi_{f}\rangle\langle\varphi_{f}\rvert$ of the final state, that projects the  state of the system to one of its possible outcomes $\lvert\varphi_{f}\rangle$. In this form we have expressed the concept of the `in-out' transition amplitude in the `in-in' language.

\subsection{Reduced Density Matrix}

The reduced density matrix $\hat{\varrho}_{\textsc{r}}$ of the subsystem we are interested in --   the atom -- can be obtained from the full density matrix $\hat{\varrho}$ of the combined system by taking the trace over the environment $E$ -- the quantum field, 
\begin{equation}\label{E:ehbdks}
	\hat{\varrho}_{\textsc{r}}=\operatorname{Tr}_{\textsc{e}}\hat{\varrho}\,,
\end{equation}
where $\operatorname{Tr}_{\textsc{e}}$ represents tracing over the environment degrees of freedom in the full density matrix operator.

In the path-integral formalism, \eqref{E:ehbdks} is given by
\begin{align}\label{E:fgrtjhd}
	\varrho_{\textsc{r}}(Q_{f}^{(\pm)};t_{f})&=\int_{-\infty}^{\infty}\!dQ_{i}^{(\pm)}\int_{Q_{i}^{(\pm)}}^{Q_{f}^{(\pm)}}\!\mathcal{D}Q^{(\pm)}\int_{-\infty}^{\infty}\!d\eta_{i}^{(\pm)}\int_{-\infty}^{\infty}\!d\eta_{f}\int_{\eta_{i}^{(\pm)}}^{\eta_{f}}\!\mathcal{D}\eta^{(\pm)}\\
	&\qquad\qquad\qquad\times\;\exp\Bigl\{i\,S[Q^{(+)},\eta^{(+)}]-i\,S[Q^{(-)},\eta^{(-)}]\Bigr\}\times\varrho(Q_{i}^{(\pm)},\eta_{i}^{(\pm)};t_{i})\,,\notag
\end{align}
where $Q$ and $\eta$ collectively represent the degrees of freedom of the reduced system and the environment, respectively. The superscript $(\pm)$ denote two time branches when we write the time evolution operator $\hat{U}$ and its Hermitian conjugate in terms of path integrations. The notations, say, $Q_{i,\,f}^{(\pm)}$, are the shorthand notation for $Q^{(\pm)}$ evaluated at $t=t_{i,\,f}$. The action $S$ of the entire system corresponds to the Hamiltonian $\hat{H}$ in \eqref{E:jbhrtr}.

If both the system and environment are Gaussian systems, and if the initial density matrix of the total system takes on a Gaussian form, the final reduced density matrix operator $\hat{\varrho}_{\textsc{r}}(t_{f})$ will remain Gaussian after we perform path integrations in \eqref{E:fgrtjhd}. It has the advantage that the coordinate representation of the reduced density matrix can be easily expressed in terms of a $2\times2$ covariance matrix $\mathcal{V}$ of the reduced system, defined by
\begin{align}\label{E:bgeirr}
	\mathcal{V}&=\operatorname{Tr}_{\textsc{s}}\Bigl(\frac{1}{2}\bigl\{\hat{R},\hat{R}^{T}\bigr\}\,\hat{\varrho}_{\textsc{r}}\Bigr)\,,&\hat{R}^{T}&=(\hat{Q},\hat{P})\,,
\end{align}
where $\hat{P}$ is the momentum operator conjugate to the system coordinate operator $\hat{Q}$ and the superscript $T$ denotes matrix transpose. That is,  $\mathcal{V}_{11}=\langle \hat{Q}^{2}\rangle$, $\mathcal{V}_{12}=\mathcal{V}_{21}=\frac{1}{2}\langle\{\hat{Q},\hat{P}\}\rangle$ and $\mathcal{V}_{22}=\langle \hat{P}^{2}\rangle$. Here we have assumed that $\langle\hat{R}\rangle=0$; otherwise we could easily replace $\hat{R}$ in \eqref{E:bgeirr} by $\hat{R}-\langle\hat{R}\rangle$. The elements of the covariance matrix can be found with the help of the Langevin equation of the reduced system~\cite{JohHu00,GalHu05,HH15}.

\subsection{Nonequilibrium dynamics}

The description of the reduced dynamics of an open system is perhaps most physically transparent by way of the Langevin equation. We will see that the internal degrees of freedom of a harmonic atom coupled to the ambient quantum field satisfy an equation of motion like that of a stochastically driven, damped oscillator.

{Following \eqref{E:fgrtjhd}, carrying out the path integrations over the environmental variables and employing the Feynman-Vernon identity~\cite{FeyVer}, we can express the remaining expression in terms of a stochastic effective action 
\begin{equation}
	\varrho_{\textsc{r}}(Q_{f}^{(\pm)};t_{f})=\int\!\mathcal{D}\xi\;P[\xi]\int_{-\infty}^{\infty}\!dQ_{i}^{(\pm)}\int_{Q_{i}^{(\pm)}}^{Q_{f}^{(\pm)}}\!\mathcal{D}Q^{(\pm)}\;\exp\Bigl\{i,S_{\textsc{eff}}[Q^{(+)},Q^{(-)};\xi]\Bigr\}
\end{equation}
if the initial density matrix $\varrho(Q_{i}^{(\pm)},\eta_{i}^{(\pm)};t_{i})$ takes a product form and the environment is Gaussian. The stochastic variable $\xi$ follows a probability distribution $P[\xi]$ determined via the Feynman-Vernon identity by the environmental dynamics and its initial state. Its physics will be clearly seen once we take the variation of the stochastic effective action $S_{\textsc{eff}}[Q^{(+)},Q^{(-)};\xi]$ to derive a Langevin equation.}

{For the goals described at the outset, we consider an atom  whose internal (electronic) degrees of freedom $Q$ are modeled by a simple harmonic oscillator  coupled to an ambient massless scalar quantum field $\phi$. The  action for this combined atom-field system takes the form
\begin{align}
	S=S_{\textsc{a}}[Q]+S_{\textsc{f}}[\phi]+S_{\textsc{int}}[Q,\phi]\,,
\end{align}
where the actions of the free harmonic oscillator and the scalar field are given respectively by
\begin{align}
	S_{\textsc{a}}[Q]&=\frac{m}{2}\,\dot{Q}^{2}-\frac{m\omega_{\textsc{b}}^{2}}{2}\,Q^{2}\,,&S_{\textsc{f}}[\phi]&=\frac{1}{2}\int\!d^{2}x\;\bigl[\partial_{\mu}\phi(x^{\mu})\bigr]^{2}\,,
\end{align}
with $\partial_{\mu}=\partial/\partial x^{\mu}$ and $x^{\mu}=(t,x)$ in a $1+1$-dimensional unbounded Minkowski spacetime. Their interaction takes the form 
\begin{equation}
	S_{\textsc{int}}[Q,\phi]=\lambda\int\!d^{2}x\;\delta^{(2)}(x^{\mu}-z^{\mu})\,\phi(x)\dot{Q}(t),
\end{equation}
where $z^{\mu}$ denotes the external (motional) spacetime position of the atom and $\lambda$ is the coupling strength between the internal degree of freedom and the field. The Langevin equation is
\begin{equation}\label{E:rivjrres}
	\ddot{Q}(t)+\omega_{\textsc{b}}^{2}Q(t)+\frac{\lambda^{2}}{m}\int^{t}_{t_{i}}\!dt'\;G_{R}^{(\phi)}(z,t;z,t')\dot{Q}(t')=-\frac{\lambda}{m}\,\dot{\xi}(t)\,,
\end{equation}
with $m$ being the mass and $\omega_{\textsc{b}}$ the bare frequency. The stochastic noise $\xi$ has the following statistical properties
\begin{align}\label{E:dbguerhsd}
	\langle\!\langle\xi(t)\rangle\!\rangle&=0\,,&\langle\!\langle\xi(t)\xi(t')\rangle\!\rangle&=G_{H}^{(\phi)}(z,t;z,t')
\end{align}
in which $G_{H}^{(\phi)}(x^{\mu},x'^{\mu})$ is the expectation value of the anti-commutator of the massless scalar field in its initial vaccum state. This and the corresponding retarded Green's function $G_{R}^{(\phi)}(x^{\mu},x'^{\mu})$ in \eqref{E:rivjrres} are consistently connected by the fluctuation-dissipation relation. They together summarize the dynamical backactions of the environment (scalar field) on the atom. The bracket $\langle\!\langle\cdots\rangle\!\rangle$ is understood as the ensemble average according to the probability distribution $P[\xi]$.}

Owing to the presence of the nonlocal expression on the left hand side of \eqref{E:rivjrres}, we see the dynamics of the oscillator is in general stochastically driven and non-Markovian in nature. For  {a single atom in} the scalar field environment, {since the retarded Green's function of the field takes the form of (the derivative of) a delta function~\cite{HH15,Greiner},} this nonlocal term will be reduced to a local damping term and frequency renormalization to $\omega_{\textsc{b}}$ 
\begin{equation}\label{E:fere}
	\ddot{Q}(t)+2\gamma\,\dot{Q}(t)+\omega^{2}Q(t)=-\frac{\lambda}{m}\,\dot{\xi}(t)\,.
\end{equation}
Here $\omega$ is the physical frequency whose actual value is determined by the experimental preparation, and $\gamma=\lambda^{2}/4 m$ is the damping constant. It is clearly seen from \eqref{E:dbguerhsd} that the stochastic variable $\xi$ acts as a driving force, its stochasticity originates from the vacuum fluctuations of the environmental quantum field. In fact, \eqref{E:fere} has the same form as the equation of motion for a charged oscillator in a quantized electromagnetic field, and the noise turns out to be the Lorentz force~\cite{HL08}. The presence of damping ushers the system to approach equilibrium for time scales greater than $\gamma^{-1}$~\cite{QTD1}. It is an attractor in this class of model in the sense that given any initial state of the oscillator, the oscillator will always end up in this equilibrium state. At strong atom-field coupling, this equilibrium state will not take a Gibbs form contrary to what is often assumed in the weak-coupling thermodynamics~\cite{QTD1}.

Our goal in this paper is to examine the transitory behavior  of this harmonic atom between different energy levels when it is strongly coupled to {an ambient} quantum field  {with  a continuous spectrum}. Thus, in an open system description, the atom will be  our system of interest and the field will act as its environment. We divide our attention between the two cases when the atom spontaneously makes a transition 1) from the ground state and 2) from an excited state. We treat the first case in the  {next section and the second case in Sec. IV}.

\section{Spontaneous Transition of an atom from its Ground State}\label{S:erbffe}


We first assume that  initially, before the atom-field interaction is turned on, the atom is in the ground state, described in the coordinate representation by
\begin{equation}\label{E:gkbdfs}
	\langle Q\vert\hat{\rho}_{\textsc{a}}^{(0)}(t_{i})\vert Q'\rangle=\rho_{\textsc{a}}^{(0)}(\Sigma_{i},\Delta_{i};t)=\biggl(\frac{m\omega}{\pi}\biggr)^{\frac{1}{2}}\,\exp\biggl\{-\frac{m\omega}{4}\Bigl[4\Sigma_{i}^{2}+\Delta_{i}^{2}\Bigr]\biggr\}\,.
\end{equation}
The variables $\Sigma_{i}$ and $\Delta_{i}$ respectively denote the center-of-mass coordinate $\Sigma=(Q+Q')/2$ and the relative coordinate $\Delta=Q-Q'$ taking their initial value at time $t_{i}$. Here the superscript $(0)$ reminds us that \eqref{E:gkbdfs} is the density matrix of the ground state for a free oscillator, and similarly the superscript $(1)$ for the first excited state. At a later time $t_{f}$, the reduced density matrix will in general take the form
\begin{equation}\label{E:gfjgs}
	\varrho_{\textsc{a}}^{(0)}(\Sigma_{f},\Delta_{f};t_{f})=\mathcal{N}\,\exp\biggl\{-a\,\Delta_{f}^{2}-i\,2b\,\Delta_{f}\Sigma_{f}-c\,\Sigma_{f}^{2}\biggr\}\,,
\end{equation}
{for the current configuration}. With the superscript $(0)$ denoting ground state, note it is important to distinguish $\hat{\varrho}_{\textsc{a}}^{(0)}(t_{f})$ from $\hat{\rho}_{\textsc{a}}^{(0)}(t_{f})$ : the former represents the reduced density matrix evaluated at time $t_{f}$ which has evolved from the ground-state density matrix \eqref{E:gkbdfs} \textit{with interaction} switched on, whereas the latter represents the density matrix \eqref{E:gkbdfs} evaluated at time $t_{f}$ of a \textit{free} oscillator.

The normalization constant $\mathcal{N}$, $a$, $b$, and $c$ can be connected with the elements of the covariance matrix for the state $\hat{\varrho}_{\textsc{a}}^{(0)}(t_{f})$ by
\begin{align}
	\mathcal{N}&=\frac{1}{\bigl[2\pi\,\langle\hat{Q}^{2}\rangle_{f}^{(0)}\bigr]^{\frac{1}{2}}}\,,\\
	a&=\frac{\langle\hat{P}^{2}\rangle_{f}^{(0)}}{2}-\frac{\bigl[\langle\{\hat{Q},\hat{P}\}\rangle_{f}^{(0)}\bigr]^{2}}{8\langle\hat{Q}^{2}\rangle_{f}^{(0)}}\,,&b&=-\frac{\langle\{\hat{Q},\hat{P}\}\rangle_{f}^{(0)}}{4\langle\hat{Q}^{2}\rangle_{f}^{(0)}}\,,&c&=\frac{1}{2\langle\hat{Q}^{2}\rangle_{f}^{(0)}}\,,\label{E:bgsjergysf}
\end{align}
where $\langle\hat{\mathcal{O}}\rangle_{f}^{(0)}$ denotes the quantum expectation value of the operator $\hat{\mathcal{O}}$ in the state described by $\hat{\varrho}_{\textsc{a}}^{(0)}(t_{f})$ in \eqref{E:gfjgs}. Similarly $\langle\mathcal{O}\rangle_{i}^{(0)}$ represents the quantum expectation value of the system operator $\hat{\mathcal{O}}$ with respect to the initial (ground-state) density matrix \eqref{E:gkbdfs}, that is,
\begin{equation}\label{E:rithersf}
	\langle\hat{\mathcal{O}}\rangle_{i}^{(0)}=\operatorname{Tr}_{Q}\Bigl\{\hat{\rho}_{\textsc{a}}^{(0)}(t_{i})\,\hat{\mathcal{O}}\Bigr\}\,,
\end{equation}
so that, for example, we have 
\begin{align}
	\mathcal{V}_{11}^{(0)}(t_{i})=\langle\hat{Q}^{2}\rangle_{i}^{(0)}&=\frac{1}{2m\omega}\,,&\mathcal{V}_{22}^{(0)}(t_{i})=\langle\hat{P}^{2}\rangle_{i}^{(0)}&=\frac{m\omega}{2}\,.
\end{align}
Thus we have explicitly expressed the reduced density matrix at $t=t_{f}$ in terms of the covariance matrix elements of the reduced system, evaluated at time $t=t_{f}$.

Next we can formally compute the transition probability of the atom from the initial ground state by \eqref{E:gksjgsx}. For example, the probability that the reduced system remains in the ground state at time $t_{f}$ will be
\begin{equation}
	P_{0\to0}=\operatorname{Tr}\Bigl\{\rho_{\textsc{a}}^{(0)}(t_{f})\,\varrho_{\textsc{a}}^{(0)}(t_{f})\Bigr\}\,.
\end{equation}
Before evaluating the transition probability, we first express the ground state density matrix \eqref{E:gkbdfs} in terms of the covariance matrix elements at the initial time,
\begin{equation}\label{E:uurkbdfs}
	\rho_{\textsc{a}}^{(0)}(\Sigma,\Delta;t)=\frac{1}{\bigl[2\pi\,\langle\hat{Q}^{2}\rangle_{i}^{(0)}\bigr]^{\frac{1}{2}}}\,\exp\biggl\{-\frac{\langle\hat{P}^{2}\rangle_{i}^{(0)}}{2}\,\Delta^{2}-\frac{1}{2\langle\hat{Q}^{2}\rangle_{i}^{(0)}}\,\Sigma^{2}\biggr\}\,.
\end{equation}
Plugging \eqref{E:uurkbdfs} into \eqref{E:gksjgsx} and carrying out the trace, we find
\begin{align}
	P_{0\to0}&=\int\!d\Sigma_{f}\,d\Delta_{f}\;\rho_{\textsc{a}}^{(0)}(\Sigma_{f},\Delta_{f};t_{f})\,\varrho_{\textsc{a}}^{(0)}(\Sigma_{f},\Delta_{f};t_{f})\notag\\
	&=\Bigl\{\Bigl[\langle\hat{Q}^{2}\rangle_{f}^{(0)}+\langle\hat{Q}^{2}\rangle_{i}^{(0)}\Bigr]\Bigl[\langle \hat{P}^{2}\rangle_{f}^{(0)}+\langle\hat{P}^{2}\rangle_{i}^{(0)}\Bigr]-\frac{1}{4}\,\Bigl[\langle\{\hat{Q},\hat{P}\}\rangle_{f}^{(0)}\Bigr]^{2}\Bigr\}^{-1}\,.\label{E:tihnsfs}
\end{align}
We can easily show that at late times $P_{0\to0}<1$. For $t\gg\gamma^{-1}$, the term $\langle\{\hat{Q},P\}\rangle_{f}\to0$ exponentially fast if the equilibrated state of the reduced system is stationary, and thus we have
\begin{align}
	&\quad\Bigl[\langle\hat{Q}^{2}\rangle_{f}^{(0)}+\langle\hat{Q}^{2}\rangle_{i}^{(0)}\Bigr]\Bigl[\langle\hat{P}^{2}\rangle_{f}^{(0)}+\langle\hat{P}^{2}\rangle_{i}^{(0)}\Bigr]-\frac{1}{4}\,\Bigl[\langle\{\hat{Q},\hat{P}\}\rangle_{f}^{(0)}\Bigr]^{2}\\
	&\simeq\Bigl[\langle\hat{Q}^{2}\rangle_{f}^{(0)}+\langle\hat{Q}^{2}\rangle_{i}^{(0)}\Bigr]\Bigl[\langle\hat{P}^{2}\rangle_{f}^{(0)}+\langle\hat{P}^{2}\rangle_{i}^{(0)}\Bigr]\geq4\sqrt{\langle\hat{Q}^{2}\rangle_{f}^{(0)}\langle\hat{Q}^{2}\rangle_{i}^{(0)}\langle\hat{P}^{2}\rangle_{f}^{(0)}\langle\hat{P}^{2}\rangle_{i}^{(0)}}\geq1\,,\notag
\end{align}
by the geometric inequality. Since in general $\hat{\varrho}_{\textsc{a}}^{(0)}(t_{f})$ is not the density matrix of the ground state, we have $\langle\hat{Q}^{2}\rangle_{f}^{(0)}\langle\hat{P}^{2}\rangle_{f}^{(0)}>1/4$. We then conclude that $0\leq P_{0\to0}<1$ at late times. Note that we did not use any perturbative expansion or any rotation-wave-like approximation to obtain the transition probability \eqref{E:tihnsfs}; so far the result is exact {for all coupling strength}.

The transition probability from the ground state to the first excited state can be computed in the same fashion. The density matrix of the first excited state of the free harmonic oscillator takes the form
\begin{align}
	\rho_{\textsc{a}}^{(1)}(\Sigma,\Delta;t)&=\biggl(\frac{m^{3}\omega^{3}}{4\pi}\biggr)^{\frac{1}{2}}\,\Bigl(4\Sigma^{2}-\Delta^{2}\Bigr)\,\exp\biggl\{-\frac{m\omega}{4}\Bigl[4\Sigma^{2}+\Delta^{2}\Bigr]\biggr\}\notag\\
	&=\frac{1}{4\sqrt{2\pi}\,\bigl[\langle\hat{Q}^{2}\rangle_{i}^{(0)}\bigr]^{\frac{3}{2}}}\,\Bigl(4\Sigma^{2}-\Delta^{2}\Bigr)\,\exp\biggl\{-\frac{\langle\hat{P}^{2}\rangle_{i}^{(0)}}{2}\,\Delta^{2}-\frac{1}{2\langle\hat{Q}^{2}\rangle_{i}^{(0)}}\,\Sigma^{2}\biggr\}\,,
\end{align}
where to facilitate the calculations, we have written the coefficients before $\Sigma$ and $\Delta$  by the elements of the covariance matrix for the initial ground state of the free oscillators, that is, those expectation values with a superscript $(0)$ and a subscript $i$, introduced in \eqref{E:rithersf},
\begin{align}
	\langle\hat{Q}^{2}\rangle_{i}^{(1)}&=3\langle\hat{Q}^{2}\rangle_{i}^{(0)}=\frac{3}{2m\omega}\,,&\langle\hat{P}^{2}\rangle_{i}^{(1)}&=\langle\hat{P}^{2}\rangle_{i}^{(0)}+\frac{1}{2\langle\hat{Q}^{2}\rangle_{i}^{(0)}}=3\langle\hat{P}^{2}\rangle_{i}^{(0)}=\frac{3m\omega}{2}\,.
\end{align}
Therefore we can show the transition probability from the ground state to the first excited state is
\begin{align}
	P_{0\to1}&=\int\!d\Sigma_{f}\,d\Delta_{f}\;\rho_{\textsc{a}}^{(1)}(\Sigma_{f},\Delta_{f};t_{f})\,\varrho_{\textsc{a}}^{(0)}(\Sigma_{f},\Delta_{f};t_{f})\notag\\
	&=\biggl\{\langle\hat{Q}^{2}\rangle_{f}^{(0)}\langle\hat{P}^{2}\rangle_{f}^{(0)}-\frac{1}{4}\,\Bigl[\langle\{\hat{Q},\hat{P}\}\rangle_{f}^{(0)}\Bigr]^{2}-\frac{1}{4}\biggr\}\label{E:thtrss}\\
	&\qquad\qquad\qquad\qquad\quad\times\biggl\{\Bigl[\langle\hat{Q}^{2}\rangle_{f}^{(0)}+\langle\hat{Q}^{2}\rangle_{i}^{(0)}\Bigr]\Bigl[\langle\hat{P}^{2}\rangle_{f}^{(0)}+\langle\hat{P}^{2}\rangle_{i}^{(0)}\Bigr]-\frac{1}{4}\,\Bigl[\langle\{\hat{Q},\hat{P}\}\rangle_{f}^{(0)}\Bigr]^{2}\biggr\}^{-\frac{3}{2}}\,.\notag
\end{align}
Since according to the Robertson-Schr\"odinger inequality~\cite{Ro29},
\begin{align}
	\langle\Delta^{2}\hat{Q}\rangle\,\langle\Delta^{2}\hat{P}\rangle&\geq\lvert\,\frac{1}{2}\bigl\{\Delta\hat{Q},\Delta\hat{P}\bigr\}\,\rvert^{2}+\lvert\,\frac{1}{2i}\langle\bigl[\Delta\hat{Q},\Delta\hat{P}\bigr]\rangle\,\rvert^{2}\,,&\Delta\hat{Q}=\hat{Q}-\langle\hat{Q}\rangle\,,
\end{align}
the denominator is semi-positive, we expect this transition probability can be {nonzero. This result is quite generic since we have not specified the initial state of the environment yet. It seems that we may always have a nonzero albeit small transition probability no matter how small it is and no matter what state the environment is initially prepared.}  In particular, this is remarkable if the environment is initially also in a non-degenerate ground state.

\subsection{Comparison with perturbation theory results}

It may be instructive to compare this result with that obtained by the time-dependent perturbation theory. The same transition probability from the ground state $\lvert E_{0}\rangle_{\textsc{a}}$ to the first excited state $\lvert E_{1}\rangle_{\textsc{a}}$, that is, the lowest two energy eigenstates, of the free atom is given by~\cite{CTQM}
\begin{align}\label{E:tturtgffs}
	P_{0\to 1}=\lambda^{2}\,\Bigl|{}_{\textsc{a}}\langle E_{1}\vert\hat{Q}(0)\vert E_{0}\rangle_{\textsc{a}}\Bigr|^{2}\,R(E_{1}-E_{0})\,,
\end{align}
where the response function $R(z)$ is defined by
\begin{equation}
	R(z)=\int_{-\infty}^{\infty}\!d\tau'\,d\tau''\,e^{-i\,z(\tau''-\tau')}\,{}_{\textsc{f}}\langle0\rvert\,\phi(\tau'')\phi(\tau')\,\rvert0\rangle_{\textsc{f}}\,,
\end{equation}
if we suppose that the environmental field is in the vacuum state $\rvert0\rangle_{\textsc{f}}$. To evaluate the response function, we first calculate the two-point function ${}_{\textsc{f}}\langle0\rvert\,\phi(t)\phi(t')\,\rvert0\rangle_{\textsc{f}}$ evaluated at a fixed spatial point, which is given by
\begin{align}
	{}_{\textsc{f}}\langle0\rvert\,\phi(t)\phi(t')\,\rvert0\rangle_{\textsc{f}}=\int_{0}^{\infty}\!\frac{d\omega}{2\pi}\;\frac{\omega}{2\pi}\,e^{-i\,\omega(t-t')}\,,
\end{align}
where $x=(t,\mathbf{x})$ and $k\cdot x=\omega\,t-\mathbf{k}\cdot\mathbf{x}$. Thus the response function $R(z)$ becomes
\begin{align}
	R(z)=\int_{0}^{\infty}\!\frac{d\omega}{2\pi}\int_{-\infty}^{\infty}\!dt\;\omega\,e^{-i\,(z+\omega) t}\,\delta(z+\omega)\,.
\end{align}
For $z>0$, the response function gives zero, so that if the ambient field is in the vacuum, then the transition probability from the ground state to the first excited state of the internal degree of freedom of the harmonic atom is zero. It means that this transition is forbidden from energy conservation considerations according to the first-order time-dependent perturbation theory.

To better understand the difference between the predictions of our nonperturbative and the traditional perturbative calculations, alternatively, we may express the density matrix \eqref{E:gfjgs} in the energy eigen-basis of the original free system Hamiltonian. That is, the element $\bigl[\hat{\varrho}_{\textsc{a}}^{(0)}(t_{f})\bigr]_{kl}={}_{\textsc{a}}\langle E_{k}\vert\,\hat{\varrho}_{\textsc{a}}^{(0)}(t_{f})\,\vert E_{l}\rangle_{\textsc{a}}$ can be written as
\begin{align}\label{E:gnkders}
	\bigl[\hat{\varrho}_{\textsc{a}}^{(0)}(t_{f})\bigr]_{kl}=\int\!dQ\,dQ'\;\psi_{k}^{*}(Q)\,\varrho_{\textsc{a}}^{(0)}(Q,Q';t_{f})\,\psi_{l}(Q')\,,
\end{align}
with $\psi_{k}(Q)=\langle Q\vert E_{k}\rangle_{\textsc{a}}$, the position representation of the energy eigenstate $\lvert E_{k}\rangle_{\textsc{a}}$. The integrals in \eqref{E:gnkders} can be found with the help of the generating function of the Hermite function $\mathcal{H}_{n}(x)$,
\begin{equation}
	e^{-s^{2}+2xs}=\sum_{n=0}^{\infty}\frac{s^{n}}{n!}\,\mathcal{H}_{n}(x)\,,
\end{equation}
and then we have
\begin{align}
	\bigl[\hat{\varrho}_{\textsc{a}}^{(0)}(t_{f})\bigr]_{kl}&=\frac{1}{\sqrt{2\pi\,\langle\hat{Q}^{2}\rangle_{f}^{(0)}}}\frac{1}{\sqrt{2^{k}k!}}\frac{1}{\sqrt{2^{l}l!}}\,\biggl(\frac{m\omega}{\pi}\biggr)^{\frac{1}{2}}\int\!dQ\,dQ'\;\exp\Bigl[-\frac{m\omega}{2}\,\bigl(Q^{2}+Q'^{2}\bigr)\Bigr]\\
	&\qquad\qquad\qquad\qquad\quad\times\mathcal{H}_{k}(\sqrt{m\omega}\,Q)\mathcal{H}_{l}(\sqrt{m\omega}\,Q')\,\exp\Bigl[-\alpha\,Q^{2}-\alpha^{*}Q'^{2}+\beta\,QQ'\Bigr]\,,\notag
\end{align}
such that
\begin{align}\label{E:rfbkhdf}
	&\quad\sum_{k,l=0}^{\infty}\frac{2^{\frac{k}{2}}r^{k}}{\sqrt{k!}}\frac{2^{\frac{l}{2}}s^{l}}{\sqrt{l!}}\,\bigl[\hat{\varrho}_{\textsc{a}}^{(0)}(t_{f})\bigr]_{kl}\notag\\
	&=\biggl[\frac{4}{(1+x)(1+y)-z^{2}}\biggr]^{\frac{1}{2}}\exp\biggl[\frac{x-y+i\,2z}{(1+x)(1+y)-z^{2}}\,r^{2}+\frac{x-y-i\,2z}{(1+x)(1+y)-z^{2}}\,s^{2}\biggr.\notag\\
	&\qquad\qquad\qquad\qquad\qquad\qquad\qquad\qquad\qquad\qquad\qquad\qquad+\biggl.\frac{2(xy-1-z^{2})}{(1+x)(1+y)-z^{2}}\,rs\biggr]\,,
\end{align}
where
\begin{align}
	\alpha&=\frac{\langle\hat{P}^{2}\rangle_{f}^{(0)}}{2}+\frac{1}{8\langle\hat{Q}^{2}\rangle_{f}^{(0)}}-\frac{\bigl[\langle\{\hat{Q},\hat{P}\}\rangle_{f}^{(0)}\bigr]^{2}}{8\langle\hat{Q}^{2}\rangle_{f}^{(0)}}-i\,\frac{\langle\{\hat{Q},\hat{P}\}\rangle_{f}^{(0)}}{4\langle\hat{Q}^{2}\rangle_{f}^{(0)}}\,,\\
	\beta&=\langle\hat{P}^{2}\rangle_{f}^{(0)}-\frac{1}{4\langle\hat{Q}^{2}\rangle_{f}^{(0)}}-\frac{\bigl[\langle\{\hat{Q},\hat{P}\}\rangle_{f}^{(0)}\bigr]^{2}}{4\langle\hat{Q}^{2}\rangle_{f}^{(0)}}\,,
\end{align}
and
\begin{align}
	x&=2m\omega\,\langle\hat{Q}^{2}\rangle_{f}^{(0)}\,,&y&=\frac{2\langle\hat{P}^{2}\rangle_{f}^{(0)}}{m\omega}\,,&z&=\langle\{\hat{Q},\hat{P}\}\rangle_{f}^{(0)}\,.
\end{align}
Eq.~\eqref{E:rfbkhdf} can be viewed as the generating function for $\bigl[\hat{\varrho}_{\textsc{a}}^{(0)}(t_{f})\bigr]_{kl}$. To find, say $\bigl[\hat{\varrho}_{\textsc{a}}^{(0)}(t_{f})\bigr]$, we merely take $n^{\text{th}}$ derivative with respect to $r$ and $m^{\text{th}}$ derivative with respect to $s$ and set $r=0=s$. The lefthand side of \eqref{E:rfbkhdf} will give
\begin{align}
	\frac{\partial}{\partial r^{n}}\frac{\partial}{\partial s^{m}}\sum_{k,l=0}^{\infty}\frac{2^{\frac{k}{2}}r^{k}}{\sqrt{k!}}\frac{2^{\frac{l}{2}}s^{l}}{\sqrt{l!}}\,\bigl[\hat{\varrho}_{\textsc{a}}^{(0)}(t_{f})\bigr]_{kl}\,\bigg|_{r=s=0}&=2^{\frac{n+m}{2}}\sqrt{n!}\sqrt{m!}\;\bigl[\hat{\varrho}_{\textsc{a}}^{(0)}(t_{f})\bigr]_{nm}\,,
\end{align}
but the righthand side in the case $n=m$ is given by
\begin{equation}
	\bigl[\hat{\varrho}_{\textsc{a}}^{(0)}(t_{f})\bigr]_{nn}=\biggl[\frac{4}{(1+x)(1+y)-z^{2}}\biggr]^{\frac{1}{2}}\bigl(b^{2}-\lvert a\rvert^{2}\bigr)^{\frac{n}{2}}\,P_{n}(\frac{b}{(b^{2}-\lvert a\rvert^{2})^{\frac{1}{2}}})
\end{equation}
with $P_{n}(z)$ the Legendre function of order $n$, and
\begin{align}
	a&=\frac{x-y+i\,2z}{(1+x)(1+y)-z^{2}}\,,&b&=\frac{xy-1-z^{2}}{(1+x)(1+y)-z^{2}}\,.
\end{align}
We can check
\begin{equation}
	\frac{\partial}{\partial r^{n}}\frac{\partial}{\partial s^{n}}\,\exp\Bigl[a\,r^{2}+a^{*}\,s^{2}+2b\,rs\Bigr]=2^{n}n!\,\bigl(b^{2}-\lvert a\rvert^{2}\bigr)^{\frac{n}{2}}\,P_{n}(\frac{b}{\sqrt{b^{2}-\lvert a\rvert^{2}}})\,.
\end{equation}
Thus we see in general $\bigl[\hat{\varrho}_{\textsc{a}}^{(0)}(t_{f})\bigr]_{nn}\neq0$; there are always some excitations because the final state of the system is not the energy eigenstate of the free-atom Hamiltonian. It is a  state which evolves out of its ground state resulting from interaction with the environment. Moreover, in general $x$, $y$ are greater than unity unless squeezing is introduced.

\subsection{Entangled state}

In fact, since $\lvert E_{0}\rangle_{\textsc{a}}\otimes\lvert0\rangle_{\textsc{f}}$ is not the eigenstate of the full Hamiltonian of the atom-field system -- this is particularly conspicuous at strong coupling -- it can not be the  lowest energy state of the combined system per se. As a matter of fact, the true ground state of the combined system is an entangled state of the free atomic state and the field state~\cite{JB04,BJ05,JB05}. Thus the product state $\lvert E_{0}\rangle_{\textsc{a}}\otimes\lvert0\rangle_{\textsc{f}}$ appears as an excited state for the combined system. The interaction between the atom and the field will then evolve the combined system from this initial product state to a resultant entangled pure state of the free atom and the scalar field. This can be best seen from the fact that the reduced system is in a mixed state, the density matrix of which has a rank greater than unity. Or alternatively, the purity of the density matrix of the reduced system is less than one. This implies the possibility of a transition of the system from the state $\lvert E_{0}\rangle_{\textsc{a}}$ to the higher state $\lvert E_{n}\rangle_{\textsc{a}}$ with $n\geq1$ while the total energy remain conserved.

We can perhaps see this argument based on atom-field entanglement more clearly when applied to a simple closed quantum system. Suppose initially at time $t_{i}$, the system is in a superposition state of the eigenstates of its Hamiltonian operator $\hat{H}$ 
\begin{align}\label{E:trndks}
	\lvert\psi_{i}\rangle=\lvert\psi(t_{i})\rangle&=\sum_{n}c_{n}(t_{i})\,\lvert\varphi_{n}\rangle\,,&\hat{H}\,\lvert\varphi_{n}\rangle&=E_{n}\,\lvert\varphi_{n}\rangle\,.
\end{align}
Initially the mean energy is given by
\begin{equation}
	\langle\psi_{i}\vert\,\hat{H}\,\vert\psi_{i}\rangle=\sum_{n}\lvert c_{n}(t_{i})\rvert^{2}\,E_{n}\,.
\end{equation}
Under unitary evolution the initial state evolves to a new state $\lvert\psi(t_{f})\rangle$ at time $t=t_{f}$, given by
\begin{align}
	\lvert\psi(t_{f})\rangle&=\hat{U}(t_{f},t_{i})\,\lvert\psi(t_{i})\rangle=\sum_{n}c_{n}(t_{f})\,\lvert\varphi_{n}\rangle=\sum_{n}c_{n}(t_{i})\,e^{-i\,E_{n}(t_{f}-t_{i})}\,\lvert\varphi_{n}\rangle\,,
\end{align}
with $\hat{U}(t_{f}-t_{i})=e^{-i\,\hat{H}\,(t_{f}-t_{i})}$. But it has the same mean energy as the initial state since the evolution is unitary
\begin{equation}
	\langle\psi_{f}\vert\,\hat{H}\,\vert\psi_{f}\rangle=\sum_{n}\lvert c_{n}(t_{f})\rvert^{2}\,E_{n}=\sum_{n}\lvert c_{n}(t_{i})\rvert^{2}\,E_{n}=\langle\psi_{i}\vert\,\hat{H}\,\vert\psi_{i}\rangle\,.
\end{equation}
However, the projection of the state $\lvert\psi_{f}\rangle$ onto any of the energy eigenstates is in general nonzero
\begin{equation}
	\langle E_{k}\vert\psi(t_{f})\rangle=c_{k}(t_{f})\neq0\,,
\end{equation}
with the probability
\begin{equation}
	P_{k}=\lvert c_{k}(t_{f})\rvert^{2}=\lvert c_{k}(t_{i})\rvert^{2}\,.
\end{equation}
Thus if we perform a projective measurement of the energy, the resulting state will be one of $\lvert\varphi_{k}\rangle$, whose corresponding energy $E_{k}$ can be greater than the mean energy $\langle \hat{H}\rangle$. One cannot use the fact that $E_{k}>\langle \hat{H}\rangle$ in this process to claim that energy is not conserved, because of one's failure to recognize large energy fluctuations in such a superposed state.

Now back to the problem under study. The product state $\lvert E_{0}\rangle_{\textsc{a}}\otimes\lvert0\rangle_{\textsc{f}}$ in principle can be expressed as a linear combination of the energy eigenstates $\lvert \mathcal{E}_{k}\rangle_{\textsc{af}}$ of the entire system
\begin{align}\label{E:rbhbgs}
	\lvert E_{0}\rangle_{\textsc{a}}\otimes\lvert0\rangle_{\textsc{f}}=\sum_{k}c_{k}\,\lvert \mathcal{E}_{k}\rangle_{\textsc{af}}.
\end{align}
Based on the argument in the previous paragraph, Eq.~\eqref{E:rbhbgs} can be interpreted as permissible transitions to the excited energy eigenstates of the entire system. Suppose both $\{\lvert\mathcal{E}_{k}\rangle_{\textsc{af}}\}$ and $\{\lvert E_{m}\rangle_{\textsc{a}}\otimes\lvert n\rangle_{\textsc{f}}\}$ are complete and non-degenerate so that there exists an inverse representation of \eqref{E:rbhbgs},
\begin{equation}
	\lvert\mathcal{E}_{k}\rangle_{\textsc{af}}=\sum_{m,n}d^{k}_{mn}\,\lvert E_{m}\rangle_{\textsc{a}}\otimes\lvert n\rangle_{\textsc{f}}\,.
\end{equation}
In particular, at the initial time $t=t_{i}$, we should have
\begin{align}
	\sum_{k}c_{k}(t_{i})\,d^{k}_{mn}(t_{i})&=\delta_{m0}\delta_{n0}\,.
\end{align}
Unitary evolution of the entire system will give
\begin{align}
	\hat{U}(t_{f},t_{i})\,\lvert E_{0}\rangle_{\textsc{a}}\otimes\lvert0\rangle_{\textsc{f}}&=\sum_{k}c_{k}(t_{i})\,e^{-i\,\mathcal{E}_{k}(t_{f}-t_{i})}\,\lvert \mathcal{E}_{k}\rangle_{\textsc{af}}=\sum_{m,n}f_{mn}(t_{f})\,\lvert E_{m}\rangle_{\textsc{a}}\otimes\lvert n\rangle_{\textsc{f}}\\
\intertext{with}
f_{mn}(t_{f})&=\sum_{k}c_{k}(t_{i})\,d^{k}_{mn}(t_{i})\,e^{-i\,\mathcal{E}_{k}(t_{f}-t_{i})}\,,
\end{align}
where the evolutionary operator $\hat{U}(t_{f},t_{i})$ takes the form
\begin{align}
	\hat{U}(t_{f},t_{i})&=e^{-i\,\hat{H}(t_{f}-t_{i})}\,,&\hat{H}&=\hat{H}_{\textsc{a}}+\hat{H}_{\textsc{f}}+\hat{H}_{\textsc{int}}\,,
\end{align}
in which $\hat{H}_{\textsc{a}}$, $\hat{H}_{\textsc{int}}$, $\hat{H}_{\textsc{f}}$ are respectively the system, interaction, field Hamiltonian. Since the phase factor $e^{-i\,\mathcal{E}_{k}(t_{f}-t_{i})}$ depends on $k$ and $t_{f}$, in general the coefficient $f_{mn}$ is not equal to $\delta_{m0}\delta_{n0}$. It implies that there exists a non-zero probability $\lvert f_{mn}(t_{f})\rvert^{2}$ that the combined system may end up in various excited states with respect to the free system and the bath.

In this section we use as example the transition from the ground state of a harmonic atom coupled to a quantum scalar field to illustrate subtle yet qualitatively important differences between results obtained from a perturbative versus a non-perturbative treatment. To begin with, 
a perturbative calculation can only describe the short time transient regime. Including higher-order contributions in a perturbative treatment often introduces secular effects, which may not be physical, or, worse yet, the resulting series may not converge. By contrast we see new physics  from a non-perturbative treatment, namely, the possibility of spontaneous excitation of an atom from its free ground state, and there is no violation of energy conservation. 
When the atom-field coupling is not vanishingly weak, the interaction term can have a comparable contribution as that of a small system. The basis spanned by the states of the free subsystem is thus not the energy eigenstates of the entire system at strong coupling. It only holds in the vanishingly weak coupling limit,
\begin{align}
	\Bigl(\hat{H}_{\textsc{a}}+\hat{H}_{\textsc{f}}+\hat{H}_{\textsc{int}}\Bigr)\lvert E_{m}\rangle_{\textsc{a}}\otimes\lvert n\rangle_{\textsc{f}}\simeq\Bigl(\hat{H}_{\textsc{a}}+\hat{H}_{\textsc{f}}\Bigr)\lvert E_{m}\rangle_{\textsc{a}}\otimes\lvert n\rangle_{\textsc{f}}=\Bigl(E_{m}+E_n^{\textsc{f}}\Bigr)\lvert E_{m}\rangle_{\textsc{a}}\otimes\lvert n\rangle_{\textsc{f}}\,,
\end{align}
where $E_n^{\textsc{f}}$ represents the energy of the field mode $\lvert n\rangle_{\textsc{f}}$. Thus, as explained earlier, in general $\lvert E_{m}\rangle_{\textsc{a}}\otimes\lvert n\rangle_{\textsc{f}}$ is the true ground state of the entire system. Recognizing this subtlety, spontaneous excitation makes sense.

We now consider transitions from the first excited state of a harmonic atom.


\section{Transition of an atom from the first excited state}\label{S:nriuthdf}

Suppose that the initial state of the internal degree of freedom is the first excited state
\begin{align}
	\varrho_{\textsc{a}}^{(1)}(\Sigma_{f},\Delta_{f};t_{f})&=\frac{1}{2\sqrt{2\pi}\bigl[\langle\hat{Q}^{2}\rangle_{f}^{(0)}\bigr]^{\frac{3}{2}}}\Bigl[\mathsf{a}\,\Delta_{f}^{2}+2\mathsf{b}\,\Delta_{f}\Sigma_{f}+\mathsf{c}\,\Sigma_{f}^{2}+\mathsf{d}\Bigr]\notag\\
	&\qquad\qquad\qquad\qquad\qquad\qquad\qquad\times\exp\biggl\{-a\,\Delta_{f}^{2}-i\,2b\,\Delta_{f}\Sigma_{f}-c\,\Sigma_{f}^{2}\biggr\}\,,
\end{align}
where
\begin{align}
	\mathsf{a}&=-\langle\hat{Q}^{2}\rangle_{f}^{(0)}\,\Bigl[\langle\hat{P}^{2}\rangle_{f}^{(1)}-\langle\hat{P}^{2}\rangle_{f}^{(0)}\Bigr]-\frac{\langle\hat{Q}^{2}\rangle_{f}^{(1)}}{4\langle\hat{Q}^{2}\rangle_{f}^{(0)}}\,\Bigl[\langle\{\hat{Q},\hat{P}\}\rangle_{f}^{(0)}\Bigr]^{2}\\
	&\qquad\qquad\qquad\qquad\qquad\qquad\qquad\qquad\qquad-\frac{\langle\{\hat{Q},\hat{P}\}\rangle_{f}^{(0)}}{4}\Bigl[\langle\{\hat{Q},\hat{P}\}\rangle_{f}^{(0)}-2\langle\{\hat{Q},\hat{P}\}\rangle_{f}^{(1)}\Bigr]\,,\notag\\
	\mathsf{b}&=\frac{i}{2}\,\langle\{\hat{Q},P\}\rangle_{f}^{(1)}-i\,\frac{\langle\hat{Q}^{2}\rangle_{f}^{(1)}}{2\langle\hat{Q}^{2}\rangle_{f}^{(0)}}\,\langle\{\hat{Q},\hat{P}\}\rangle_{f}^{(0)}\,,\qquad\qquad\mathsf{c}=\frac{\langle\hat{Q}^{2}\rangle_{f}^{(1)}}{\langle\hat{Q}^{2}\rangle_{f}^{(0)}}-1\,,\\
	\mathsf{d}&=-\langle\hat{Q}^{2}\rangle_{f}^{(1)}+3\langle\hat{Q}^{2}\rangle_{f}^{(0)}\,.
\end{align}
We can double check the results by taking the limit $t_{f}\to t_{i}$ where
\begin{align*}
	\langle\hat{Q}^{2}\rangle_{f}^{(0)}&\to\frac{1}{2m\omega}\,,&\langle\hat{Q}^{2}\rangle_{f}^{(1)}&\to3\langle\hat{Q}^{2}\rangle_{f}^{(0)}\,,&\langle\hat{P}^{2}\rangle_{f}^{(0)}&\to\frac{m\omega}{2}\,,&\langle\hat{P}^{2}\rangle_{f}^{(1)}&\to3\langle\hat{P}^{2}\rangle_{f}^{(0)}\,,\\
	\langle\{\hat{Q},\hat{P}\}\rangle_{f}^{(0)}&\to0\,,&\langle\{\hat{Q},\hat{P}\}\rangle_{f}^{(1)}&\to1\,,
\end{align*}	
and we find that $\hat{\varrho}_{\textsc{a}}^{(1)}(t_{f})\to\hat{\rho}_{\textsc{a}}^{(1)}(t_{i})$.

The transition probability $P_{1\to0}$ from the first excited state to the ground state is then straightforwardly given by
\begin{align}
	P_{1\to0}&=\int\!d\Sigma_{f}\,d\Delta_{f}\;\rho_{\textsc{a}}^{(0)}(\Sigma_{f},\Delta_{f};t_{f})\,\varrho_{\textsc{a}}^{(1)}(\Sigma_{f},\Delta_{f};t_{f})\notag\\
	&=\biggl\{2\langle\hat{Q}^{2}\rangle_{f}^{(0)}\langle\hat{P}^{2}\rangle_{f}^{(0)}-\frac{1}{2}\,\Bigl[\langle\{\hat{Q},\hat{P}\}\rangle_{f}^{(0)}\Bigr]^{2}+\frac{1}{4}+\frac{1}{4}\,\langle\{\hat{Q},\hat{P}\}\rangle_{f}^{(0)}\langle\{\hat{Q},\hat{P}\}\rangle_{f}^{(1)}-\frac{1}{2}\,\langle\hat{Q}^{2}\rangle_{f}^{(0)}\langle\hat{P}^{2}\rangle_{f}^{(1)}\biggr.\notag\\
	&\qquad-\biggl.\frac{1}{2}\,\langle\hat{Q}^{2}\rangle_{f}^{(1)}\langle\hat{P}^{2}\rangle_{f}^{(0)}-\frac{1}{2}\,\langle\hat{Q}^{2}\rangle_{i}^{(0)}\Bigl[\langle\hat{P}^{2}\rangle_{f}^{(1)}-3\langle\hat{P}^{2}\rangle_{f}^{(0)}\Bigr]-\frac{1}{2}\,\langle\hat{P}^{2}\rangle_{i}^{(0)}\Bigl[\langle\hat{Q}^{2}\rangle_{f}^{(1)}-3\langle\hat{Q}^{2}\rangle_{f}^{(0)}\Bigr]\biggr\}\notag\\
	&\times\biggl\{\Bigl[\langle\hat{Q}^{2}\rangle_{f}^{(0)}+\langle\hat{Q}^{2}\rangle_{i}^{(0)}\Bigr]\Bigl[\langle\hat{P}^{2}\rangle_{f}^{(0)}+\langle\hat{P}^{2}\rangle_{i}^{(0)}\Bigr]-\frac{1}{4}\,\Bigl[\langle\{\hat{Q},\hat{P}\}\rangle_{f}^{(0)}\Bigr]^{2}\biggr\}^{-\frac{3}{2}}\,.\label{E:thgbdg}
\end{align}
We can perform a consistency check of this result. We have argued that when the dynamics of the reduced system is fully relaxed, the covariant matrix elements are independent of the initial state. Let us assume,  in the weak atom-field coupling limit, that their values at late times are a little greater than the corresponding initial values, so we can make such assignments
\begin{align}
	\langle\hat{Q}^{2}\rangle_{f}^{(1)}&\simeq\langle\hat{Q}^{2}\rangle_{f}^{(0)}\simeq\langle\hat{Q}^{2}\rangle_{i}^{(0)}\bigl(1+2\delta\bigr)\,,&\langle\hat{P}^{2}\rangle_{f}^{(1)}&\simeq\langle\hat{P}^{2}\rangle_{f}^{(0)}\simeq\langle\hat{P}^{2}\rangle_{i}^{(0)}\bigl(1+2\epsilon\bigr)\,,\\
	\langle\{\hat{Q},\hat{P}\}\rangle_{f}^{(1)}&\simeq0\,,&\langle\{\hat{Q},\hat{P}\}\rangle_{f}^{(0)}&\simeq0\,,
\end{align}
with $\delta$, $\epsilon$ being small positive numbers. The transition probability is then given approximately by
\begin{equation}
	P_{1\to0}\simeq\frac{1}{\sqrt{\bigl(1+\delta\bigr)\bigl(1+\epsilon\bigr)}}\lesssim1\,.
\end{equation}
This is reasonable. Meanwhile, we may estimate the transition probabilities found earlier in \eqref{E:tihnsfs} and \eqref{E:thtrss}
\begin{align}
	P_{0\to0}&\simeq\frac{1}{\sqrt{\bigl(1+\delta\bigr)\bigl(1+\epsilon\bigr)}}\simeq 1-\frac{\delta+\epsilon}{2}+\mathcal{O}(\epsilon^{2}\wedge\delta^{2}\wedge\epsilon\delta)\,,\\
	P_{0\to1}&\simeq\frac{\delta+\epsilon+\delta\epsilon}{2\bigl[\bigl(1+\delta\bigr)\bigl(1+\epsilon\bigr)]^{\frac{3}{2}}}\simeq\frac{\delta+\epsilon}{2}+\mathcal{O}(\epsilon^{2}\wedge\delta^{2}\wedge\epsilon\delta)\,.
\end{align}
Note that the sum of these two probabilities is still smaller than unity,
\begin{equation}
	P_{0\to0}+P_{0\to1}\simeq1-\frac{3\delta^{2}+2\delta\epsilon+3\epsilon^{2}}{8}+\mathcal{O}(\epsilon^{3}\wedge\delta^{3}\wedge\epsilon^{2}\delta\wedge\epsilon\delta^{2})\,.
\end{equation}
The contributions from the transitions to higher levels, even though nonzero, are one order of magnitude smaller in the weak coupling limit.

We can also further compute the probability $P_{1\to1}$ that the system remains in the first excited state and the probability $P_{1\to2}$ of it leaping to the second excited state. However, even though these calculations are nothing but Gaussian integrals, the obtained transient expressions are too massive. So we will only show the transition probabilities in the long time limit, that is, the equilibrium configuration after the motion of the reduced system is fully relaxed. The asymptotic probability $P_{1\to1}(\infty)$ is given by
\begin{align}
	P_{1\to1}(\infty)&=\bigg\{\langle\hat{Q}^{2}\rangle_{f}^{(0)}\langle\hat{P}^{2}\rangle_{f}^{(0)}-\frac{1}{4}\biggr\}\times\biggl\{\Bigl[\langle\hat{Q}^{2}\rangle_{f}^{(0)}+\langle\hat{Q}^{2}\rangle_{i}^{(0)}\Bigr]\Bigl[\langle\hat{P}^{2}\rangle_{f}^{(0)}+\langle\hat{P}^{2}\rangle_{i}^{(0)}\Bigr]\biggr\}^{-\frac{3}{2}}\,,\label{E:jgbkssd}\\
\intertext{and $P_{1\to2}$ by}
	P_{1\to2}(\infty)&=\biggl\{\frac{1}{2}\Bigl[\langle\hat{Q}^{2}\rangle_{f}^{(0)}\langle\hat{P}^{2}\rangle_{i}^{(0)}-\langle\hat{Q}^{2}\rangle_{i}^{(0)}\langle\hat{P}^{2}\rangle_{f}^{(0)}\Bigr]^{2}+\Bigl[\langle\hat{Q}^{2}\rangle_{f}^{(0)}\langle\hat{P}^{2}\rangle_{f}^{(0)}-\frac{1}{4}\Bigr]^{2}\biggr\}\notag\\
	&\qquad\qquad\qquad\qquad\qquad\times\biggl\{\Bigl[\langle\hat{Q}^{2}\rangle_{f}^{(0)}+\langle\hat{Q}^{2}\rangle_{i}^{(0)}\Bigr]\Bigl[\langle\hat{P}^{2}\rangle_{f}^{(0)}+\langle\hat{P}^{2}\rangle_{i}^{(0)}\Bigr]\biggr\}^{-\frac{5}{2}}\,.
\end{align}
Under the weak coupling assumption we can give the following estimations
\begin{align}
	P_{1\to1}(\infty)&\simeq\frac{\delta+\epsilon+\delta\epsilon}{2\bigl[\bigl(1+\delta\bigr)\bigl(1+\epsilon\bigr)]^{\frac{3}{2}}}\simeq\frac{\delta+\epsilon}{2}+\mathcal{O}(\epsilon^{2}\wedge\delta^{2}\wedge\epsilon\delta)\,,\\
	P_{1\to2}(\infty)&\simeq\frac{3\delta^{2}+2\delta\,\epsilon+3\epsilon^{2}+8\delta\epsilon^{2}+8\delta^{2}\epsilon+8\delta^{2}\epsilon^{2}}{8\bigl[\bigl(1+\delta\bigr)\bigl(1+\epsilon\bigr)]^{\frac{5}{2}}}\simeq\frac{3\delta^{2}+2\delta\epsilon+3\epsilon^{2}}{8}+\mathcal{O}(\epsilon^{3}\wedge\delta^{3}\wedge\epsilon^{2}\delta\wedge\epsilon\delta^{2})\,.\label{E:gkfresdf}
\end{align}
The sum of $P_{1\to0}$, $P_{1\to1}$ and $P_{1\to2}$ is
\begin{equation}
	P_{1\to0}+P_{1\to1}+P_{1\to2}=1-\frac{5\delta^{3}+3\delta^{2}\epsilon+3\delta\epsilon^{2}+5\epsilon^{3}}{16}+\cdots\,.
\end{equation}
Again, it shows the possibility of spontaneous excitation in terms of the free atomic states; however, in the weak atom-field coupling limit the transitions to higher levels becomes less and less likely because of higher-order dependence on the coupling constant. This is compatible with the predictions by the perturbative treatment. On the other hand, at strong coupling limit, these ``forbidden transitions'', although relatively smaller, are not vanishingly small.

\section{Examples, Analysis and Comparison}\label{S:rtkbgfhgse}

Earlier we have presented the exact results for the transition probability of a harmonic atom  strongly coupled to a free quantum scalar field. 
In this section we take these formal expressions and show explicitly the transition probability from the first excited state to the ground state $P_{1\to0}$ via \eqref{E:thgbdg}, and give a quantitative comparison with that derived from time-dependent perturbation theory.  Since at strong atom-field coupling  there will be more than one scale that have comparable magnitudes  the transient dynamics of the transition probability is expected to be more complicated than that in time-dependent perturbation theory. 

\subsection{Late time saturation behavior}

We first examine the late-time $(\gamma t\gg1)$ result. To do so, based on \eqref{E:thgbdg}, we need the late-time expressions for the elements of the covariance matrix. From numerous model studies of the dynamics of oscillator-field systems~\cite{CalHu}, we know that at late times  \textit{the behavior of the reduced system is governed by the environment}. Thus we have~\cite{LinHu06,HH15,QTD1}
\begin{align}
	\lim_{t\to\infty}\langle\hat{Q}^{2}(t)\rangle_{f}^{(1)}&=\lim_{t\to\infty}\langle\hat{Q}^{2}(t)\rangle_{f}^{(0)}=\frac{2}{m}\operatorname{Im}\int_{0}^{\infty}\!\frac{d\kappa}{2\pi}\;\tilde{d}_{2}(\kappa)\,,\label{E:gbkshgb1}\\
	\lim_{t\to\infty}\langle\hat{P}^{2}(t)\rangle_{f}^{(1)}&=\lim_{t\to\infty}\langle\hat{P}^{2}(t)\rangle_{f}^{(0)}=2m\operatorname{Im}\int_{0}^{\infty}\!\frac{d\kappa}{2\pi}\;\kappa^{2}\tilde{d}_{2}(\kappa)\,,\label{E:gbkshgb2}\\
	\lim_{t\to\infty}\langle\bigl\{\hat{Q}(t),\hat{P}(t)\bigr\}\rangle_{f}^{(1)}&=\lim_{t\to\infty}\langle\bigl\{\hat{Q}(t),\hat{P}(t)\bigr\}\rangle_{f}^{(0)}=-i\,2\operatorname{Im}\int_{0}^{\infty}\!\frac{d\kappa}{2\pi}\;\kappa\,\tilde{d}_{2}(\kappa)\,,\label{E:gbkshgb3}
\end{align}
where $\tilde{d}_{2}(\kappa)$ takes the form
\begin{equation}
	\tilde{d}_{2}(\kappa)=\frac{1}{-\kappa^{2}+\omega^{2}-\dfrac{\lambda^{2}}{m}\,\tilde{G}_{R}^{(\phi)}(\kappa)}\,,
\end{equation}
and in fact can be identified as the retarded Green's function for the internal degree of freedom of the harmonic atom with respect to the Langevin equation \eqref{E:rivjrres}. The function $\tilde{G}_{R}^{(\phi)}(\kappa)$ is the Fourier transform of the retarded Green's function of the scalar field, defined by
\begin{equation}
	\tilde{G}_{R}^{(\phi)}(\kappa)=\int_{-\infty}^{\infty}\!d\tau\;G_{R}^{(\phi)}(\tau)\,e^{+i\kappa\tau}\,.
\end{equation}
In particular, Eq.~\eqref{E:gbkshgb3} gives zero. This is easily seen when we write \eqref{E:gbkshgb3} in  the form
\begin{align}
	\lim_{t\to\infty}\langle\bigl\{\hat{Q}(t),\hat{P}(t)\bigr\}\rangle_{f}^{(0)}&=-i\,\frac{e^{2}}{m}\int_{-\infty}^{\infty}\!\frac{d\kappa}{2\pi}\;\kappa\,\lvert\tilde{d}_{2}(\kappa)\rvert^{2}\,\tilde{G}_{H}^{(\phi)}(\kappa)\,.
\end{align}
The integrand in this case is an odd function of $\kappa$, so the integral over $\omega$ will vanish. We have assumed that the field is initially in its vacuum state, so the corresponding Hadamard function $G_{H}^{(\phi)}(t,t')$ of the field at the spatial location of the atom is given by
\begin{align}
	G_{H}^{(\phi)}(t-t')=\frac{1}{2}\,\langle\bigl\{\phi(x^{\mu}),\phi(x'^{\mu})\bigr\}\rangle=\int_{-\infty}^{\infty}\!\frac{d\kappa}{2\pi}\;\operatorname{sgn}(\kappa)\,\frac{\kappa}{4\pi}\,e^{-i\kappa(t-t')}\,,
\end{align}
and $\tilde{G}_{H}^{(\phi)}(\kappa)$ is its Fourier transformation. Thus the transition probability from the excited to the ground state is greatly reduced to
\begin{align}
	P_{1\to0}&=\biggl\{\Bigl[\langle\hat{Q}^{2}\rangle_{f}^{(0)}+\langle\hat{Q}^{2}\rangle_{i}^{(0)}\Bigr]\Bigl[\langle\hat{P}^{2}\rangle_{f}^{(0)}+\langle\hat{P}^{2}\rangle_{i}^{(0)}\Bigr]\biggr\}^{-\frac{1}{2}}\,,\label{E:hnlhdklrt}
\end{align}
with $\langle\hat{Q}^{2}\rangle_{i}^{(0)}=1/(2m\omega)$, $\langle\hat{P}^{2}\rangle_{i}^{(0)}=m\omega/2$, where $\omega$ is  the physical frequency of the internal degree of freedom of the atom. {As has been argued earlier, $P_{1\to0}$ in Eq.~\eqref{E:hnlhdklrt} can  possibly be less than unity.}

Let us now look at time-dependent perturbation theory (TDPT). Since it applies only when the interaction between the atom and the field is sufficiently weak, one does not need to consider modification of the wave function of the atom's internal degree freedom. Thus at the final time, the covariance matrix elements will still take on the \textit{same values as at the initial time}, namely,
\begin{align}\label{E:gnkdgskd}
	\langle\hat{Q}^{2}\rangle_{f}^{(0)}&=\frac{1}{2m\omega}\,,&\langle\hat{P}^{2}\rangle_{f}^{(0)}&=\frac{m\omega}{2}\,.
\end{align}
This implies that the transition probability is unity, that is, in the field vacuum, all excited atoms will eventually fall to the ground state without exception 
\begin{align}\label{E:oeirhbdf}
	\lim_{\gamma\to0}P_{1\to0}&=1\,.
\end{align}
Note {to obtain} \eqref{E:oeirhbdf} one should first take the $\gamma\to0$ limit -- {because perturbation theory functions only under the condition of vanishingly weak coupling} -- before taking the late time limit $t\to\infty$ in \eqref{E:gnkdgskd}. These are the tacit assumptions behind the TDPT of spontaneous emission.  As is well-known one can also phrase this in terms of Einstein's  $A$, $B$ coefficients. Phenomenologically, for a simple two-level atom in the field vacuum, the change of the number density $n_{e}$ of the atoms in the excited state per unit time is given by
\begin{align}\label{E:gkhshfd}
	\frac{dn_{e}}{dt}&=-A\,n_{e}\,,&A&>0\,,
\end{align}
where $A$ is the Einstein coefficient for spontaneous emission. (Its actual expression is not important for the argument.) It is easily seen that the number density $n_{e}$ of the atoms in the excited state exhibits exponential decay in time
\begin{equation}\label{E:gvjhyerd}
	n_{e}(t)=n_{2}(0)\,e^{-At}\,.
\end{equation}
If initially all of the atoms are in the exited state, we have $n_{e}(0)=n$, where $n$ is the number density of the atoms, and as $t$ approach infinity we find $n_{e}$ goes to zero exponentially fast, that is, $n_{e}(\infty)=0$. Thus we may conclude the number density $n_{g}=n-n_{e}$ of the atoms in the ground state will approach $n$, or the ratio $n_{g}/n\to1$ in the end.

When the interaction between the atom and the field is \textit{not }infinitesimally small, we cannot ignore the change in the wave function amplitude of the atom, namely, the back-action effect,  manifested as dissipation. In addition, the cutoff scale in the theory cannot be simply hand-waved away. As will be seen later, the cutoff scale $\Lambda$ in the current case usually appears inside the logarithm and is paired with the damping constant $\gamma$ in the form $\sim\gamma\,\ln\Lambda$. This is the reason why one can ignore it when $\gamma$ is vanishingly small but not so when $\gamma$ is finite. Therefore, the covariance matrix elements in general will not take on the simple forms given by \eqref{E:gnkdgskd}, as there are additional scales $\gamma$ and $\Lambda$ present.  By \eqref{E:gbkshgb1} and \eqref{E:gbkshgb2}, we find
\begin{align}
	\langle\hat{Q}^{2}\rangle_{f}^{(0)}&=\frac{1}{2m\Omega}-\frac{1}{\pi m\Omega}\,\tan^{-1}\frac{\gamma}{\Omega}\,,\\
	\langle\hat{P}^{2}\rangle_{f}^{(0)}&=\frac{m(\Omega^{2}-\gamma^{2})}{2\Omega}\biggl\{1-\frac{1}{\pi}\bigg[\tan^{-1}\frac{\gamma}{\Lambda-\Omega}-\tan^{-1}\frac{\gamma}{\Lambda+\Omega}+2\tan^{-1}\frac{\gamma}{\Omega}\biggr]\biggr\}\\
	&\qquad\qquad\qquad\qquad\qquad\qquad\qquad\qquad+\frac{m\gamma}{2\pi}\ln\frac{[(\Lambda-\Omega)^{2}+\gamma^{2}][(\Lambda+\Omega)^{2}+\gamma^{2}]}{(\Omega^{2}+\gamma^{2})^{2}}\,,\notag
\end{align}
where $\Omega=\sqrt{\omega^{2}-\gamma^{2}}$ is the resonance frequency and $\gamma=\lambda^{2}/4 m$ is the damping constant. They reduce to \eqref{E:gnkdgskd} in the limit $\gamma\to0$,
\begin{align}
	\langle\hat{Q}^{2}\rangle_{f}^{(0)}&\to\frac{1}{2m\omega}-\frac{\gamma}{\pi m\omega^{2}}+\mathcal{O}(\gamma^{2})\,,\label{E:gkjfgds}\\
	\langle\hat{P}^{2}\rangle_{f}^{(0)}&\to\frac{m\omega}{2}+\frac{m\gamma}{\pi}\biggl\{\ln\Bigl[\frac{\Lambda^{2}}{\omega^{2}}-1\Bigr]-\Big(1-\frac{\omega^{2}}{\Lambda^{2}}\Bigr)^{-1}\biggr\}+\mathcal{O}(\gamma^{2})\,.\label{E:rtbrhvg}
\end{align}
\begin{figure}
\centering
    \scalebox{0.6}{\includegraphics{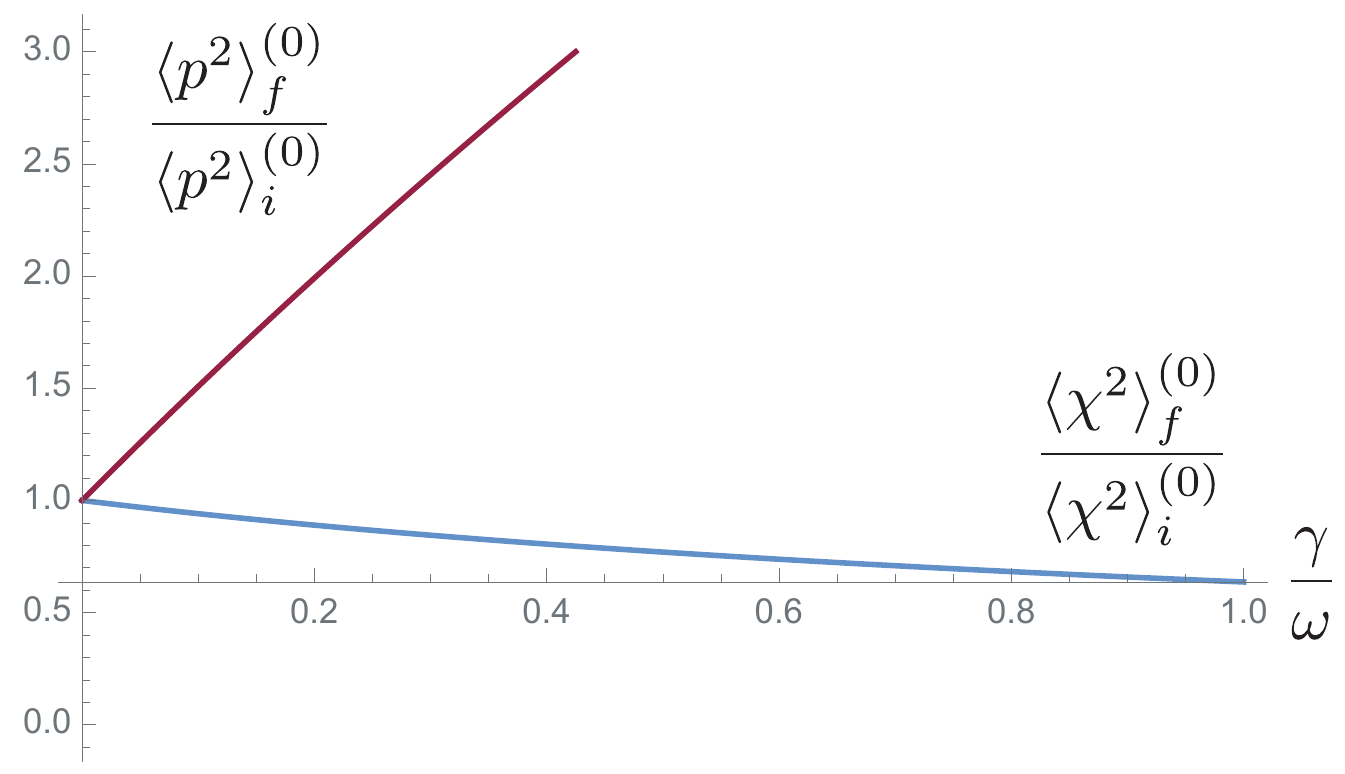}}
    \caption{Effects of atom-field coupling strength on the elements of covariance matrix. The cutoff scale is chosen such that $\Lambda/\omega=100$. We see that the expectation value of the momentum is more dramatically affected due to the non-negligible presence of the energy cutoff scale, as seen in \eqref{E:rtbrhvg}.}\label{Fi:strcpling}
\end{figure}

As seen from Fig.~\ref{Fi:strcpling}, at strong coupling, the values of the covariance matrix elements can significantly differ from those in the weak coupling limit. In particular, the cutoff scale can be important at strong coupling for $\langle\hat{P}^{2}\rangle$. Understandably these strong-coupling effects will manifest in the transition probability \eqref{E:hnlhdklrt}. At late times $t\gg\gamma^{-1}$, we find that the transition probability from the first excited state to the ground state is approximately given by
\begin{equation}\label{E:tjkwndsfs}
	P_{1\to0}(\infty)\simeq1+\frac{\gamma}{2\pi\omega\bigl(\Lambda^{2}-\omega^{2}\bigr)}\biggl[2\Lambda^{2}-\omega^{2}-\bigl(\Lambda^{2}-\omega^{2}\bigr)\ln\frac{\Lambda^{2}-\omega^{2}}{\omega^{2}}\biggr]+\mathcal{O}(\gamma^{2})\,.
\end{equation}
The finite $\gamma$ correction contains a cutoff-dependent expression in the form $\gamma\,\ln\Lambda$, so it changes  mildly with the cutoff scale. In addition, we note that the transition probability gradually deviates from, but remains less than, unity when the damping constant $\gamma$ increases from zero. The difference can be quite noticeable when $\gamma$ is no longer small. This also implies that the atom in the excited state does not always emit field quanta and falls to the ground state. This can be understood according to the earlier discussion that the product state the atom initially resides in is not necessarily the energy-eigenstate  of the full system,  their difference increases with the atom-field interaction strength. 
\begin{figure}
\centering
    \scalebox{0.45}{{\huge(a)}\includegraphics{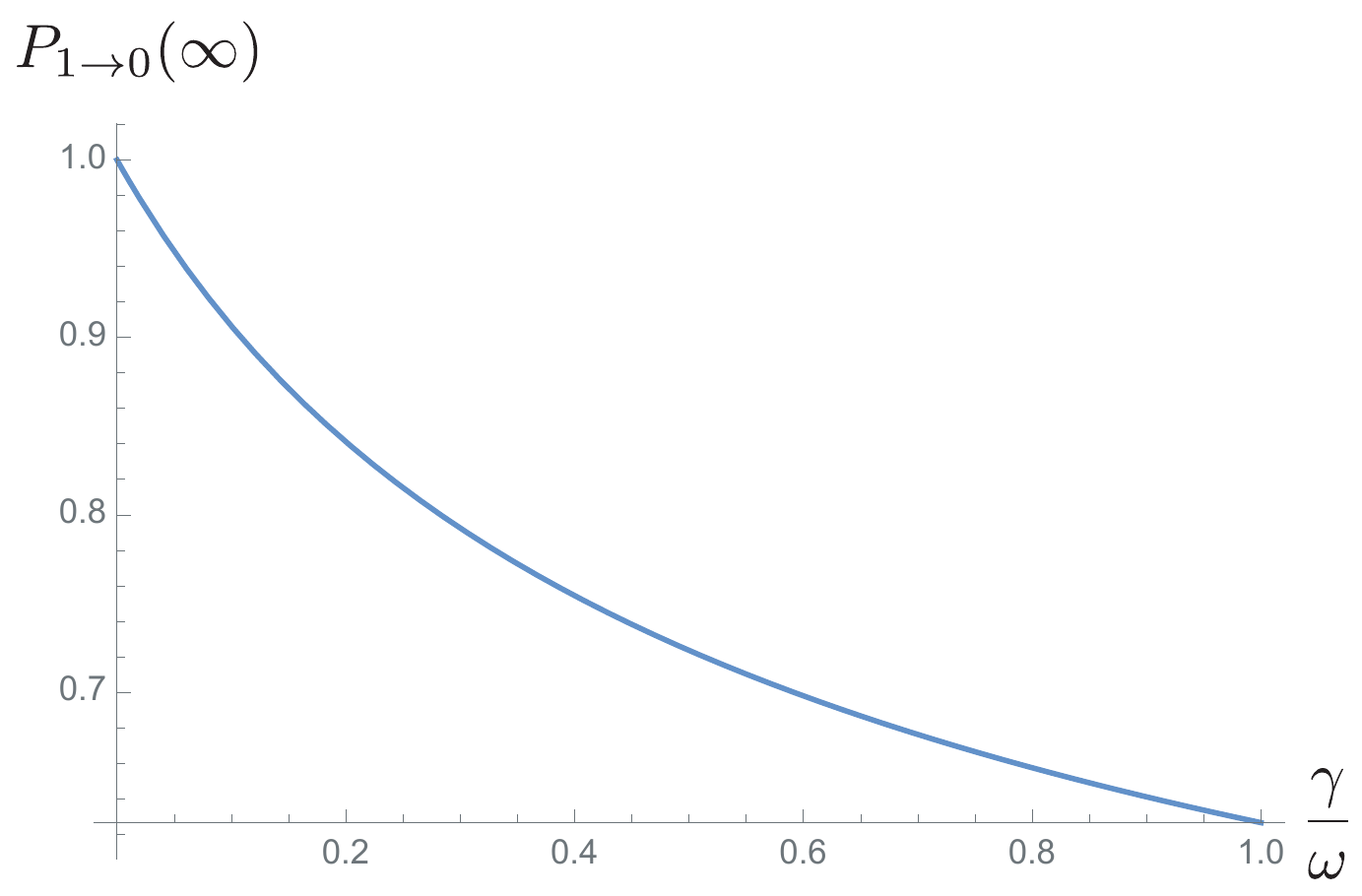}{\huge(b)}\includegraphics{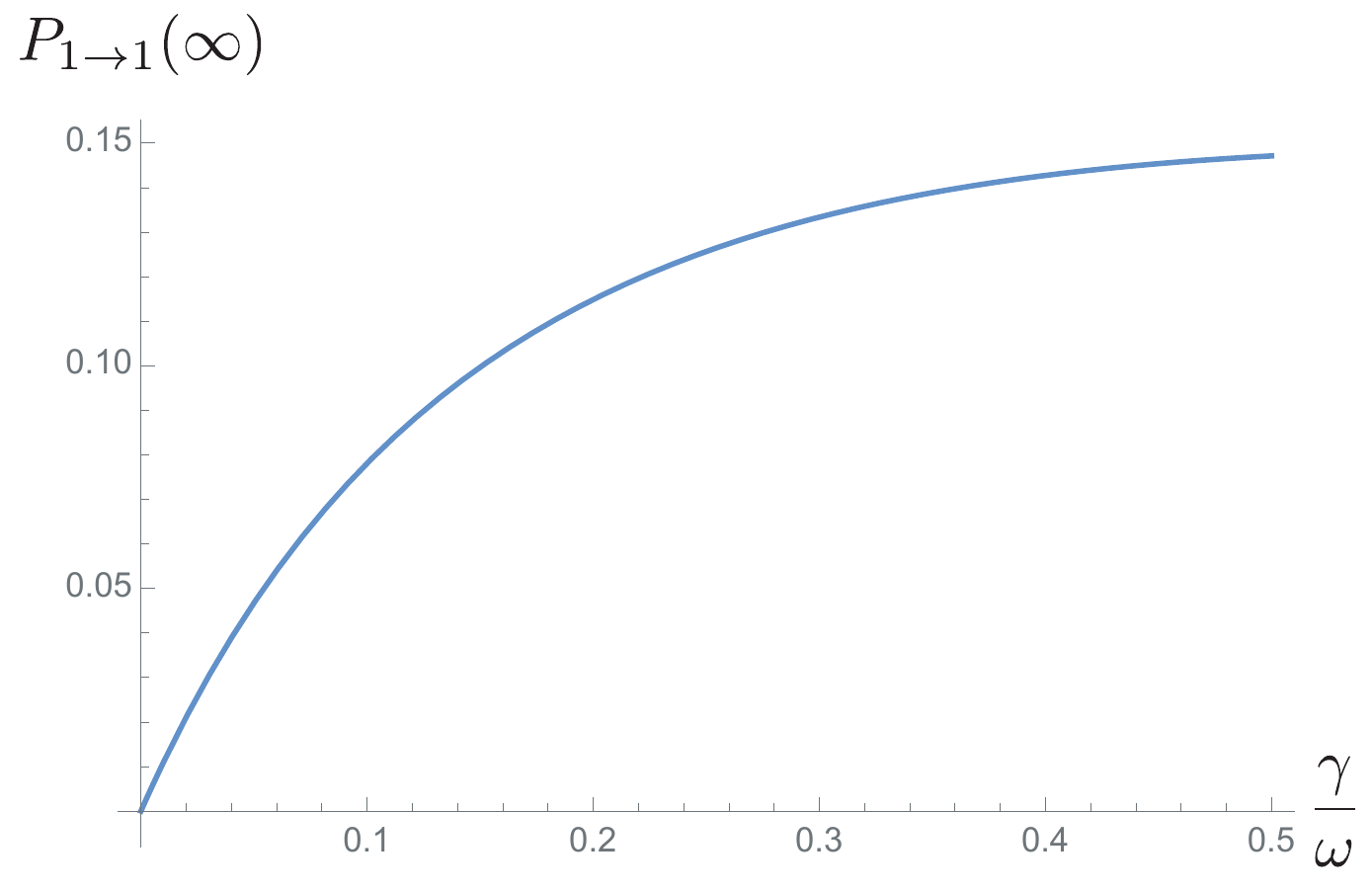}}
    \caption{To illustrate the effects of  strong atom-field interaction, we plot the transition probabilities of the harmonic atom, initially in the first excited state, versus the damping constant $\gamma$ when it makes (a) a transition to the ground state, and (b) remains in the first excited state. }\label{Fi:transPro}
\end{figure}

We note that there is a finite probability that the atom in the first excited state may sneak up to the higher excited states. This can also be understood in terms of the density of the state of the harmonic atom. Before the atom-field interaction is switched on, the harmonic atom is a gapped system and its density of states take on the form $\delta(E-(n+1/2)\omega)$. However, when the interaction is turned on, after the system reaches equilibrium, the atom becomes gapless and has a continuous distribution of states if the interaction strength takes a finite value~\cite{HZ95}. The delta-function form of the density of state will become Breit-Wigner peaks of finite width when interaction strength is still weak, but these prominent features will gradually disappear when the peak widths broaden with increasing coupling strength. The excited atom is thus not always destined to transit to the ground state.

For example, from \eqref{E:jgbkssd}, we find that the probability that the atom remains in the first excited state is nonzero, and takes the form
\begin{equation}
	P_{1\to1}\simeq-\frac{\gamma}{2\pi\omega\bigl(\Lambda^{2}-\omega^{2}\bigr)}\biggl[2\Lambda^{2}-\omega^{2}-\bigl(\Lambda^{2}-\omega^{2}\bigr)\ln\frac{\Lambda^{2}-\omega^{2}}{\omega^{2}}\biggr]+\mathcal{O}(\gamma^{2})\,,
\end{equation}
which is positive, and the transition to the second excited state will have the probability of order $\mathcal{O}(\gamma^{2})$, as has been qualitatively discussed in \eqref{E:gkfresdf}.

\begin{figure}
	\centering
    \scalebox{0.5}{\includegraphics{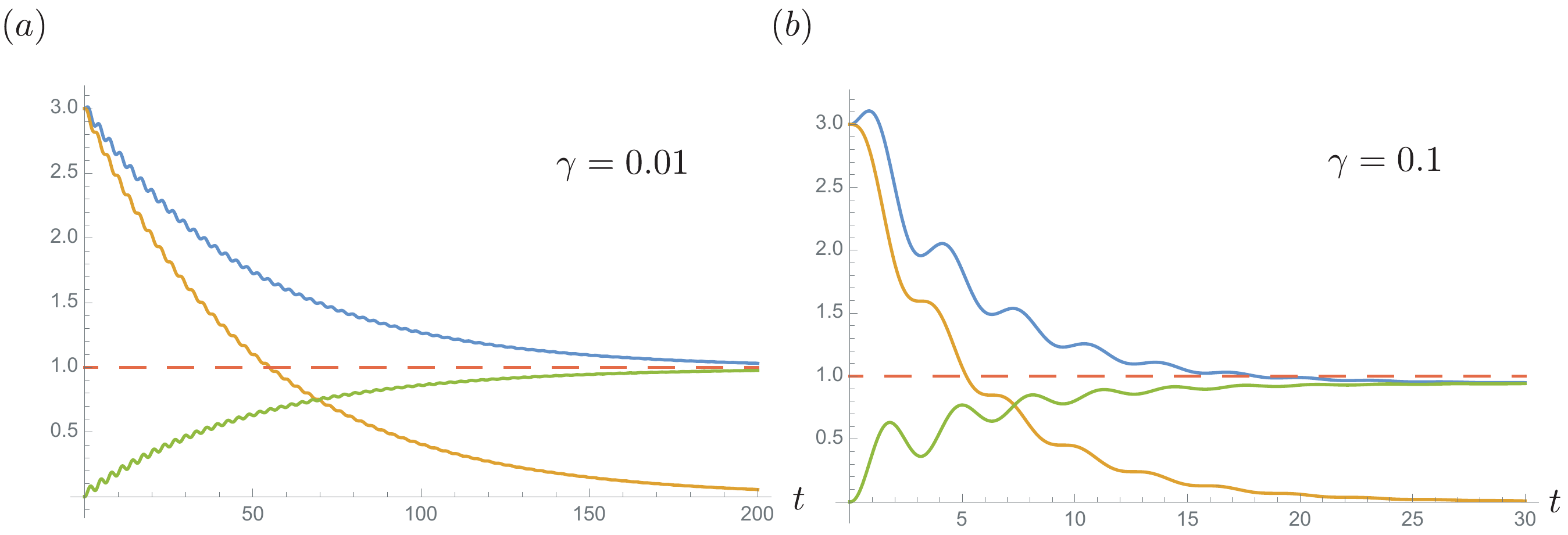}}
    \caption{We plot, for two different values of the damping constant, the time evolution of $\langle\hat{Q}^{2}(t)\rangle$, normalized by {the corresponding value  of a free oscillator in the ground state}. We fix the cutoff frequency and oscillator frequency for $\Lambda=100$, $\omega=1$. The orange and the green curves represent the contribution from the intrinsic and the induced components, respectively. The blue curve describes the total contribution in \eqref{E:nhgjkhd}. We see in (b) that when the damping constant is comparable to the oscillator frequency, the transient transitions start showing a more prominent oscillatory feature than in (a).}\label{Fi:x2}
\end{figure}


\subsection{Early time transient behavior}

The actual transient behavior of the transition probability calculated without making the weak coupling assumption is more complex than that obtained from time-dependent perturbation theory in that there exist additional time scales such as $\gamma^{-1}$, $\Lambda^{-1}$ and possibly the distance $d$ to the boundary when  dielectric substance is present. Here, we consider the underdamped case which typically has $\Lambda^{-1}<\omega^{-1}<\gamma^{-1}$. The overdamped case tends to induce instability if non-Markovian effects are non-negligible, for example, when there exists spatial boundary or when the inter-atomic coupling is included in the consideration~\cite{AFM1}. The transient dynamics of the transition probability is best understood from inspecting the general features of the dynamics of the covariance matrix elements of the harmonic atom. Consider for example the displacement uncertainty $\langle\hat{Q}^{2}(t)\rangle_{f}^{(1)}$ of the atom's internal degrees of freedom in an unbounded space, 
\begin{equation}\label{E:nhgjkhd}
	\langle\hat{Q}^{2}\rangle_{f}^{(1)}=d_{1}^{2}(t_{f})\,\langle\hat{Q}^{2}\rangle_{i}^{(1)}+d_{2}^{2}(t_{f})\,\frac{\langle\hat{P}^{2}\rangle_{i}^{(1)}}{m^{2}}+\frac{e^{2}}{m^{2}}\int^{t_{f}}_{0}\!ds\!\int^{t_{f}}_{0}\!ds'\;d_{2}(s)d_{2}(s')\,G_{H}^{(\phi)}(s,s')\,,
\end{equation}
when the atom is initially in the first excited state. The first two terms in \eqref{E:nhgjkhd} is the intrinsic or active component of the displacement uncertainty because it tells the quantum fluctuations of the atom when the atom-field interaction is absent. The last term in \eqref{E:nhgjkhd} is the induced or passive component because it contains the  quantum fluctuations acquired from the environmental field after the interaction is switched on. They have very distinct temporal evolutions. The intrinsic component exponentially decays with a time scale $(2\gamma)^{-1}$ due to dissipative back-action from the field, while the induced component initially grows linearly with time driven by the quantum noise from the environment field, but later, after $t\gg\gamma^{-1}$,  saturates to a time-independent value after the friction force picks up. The existence of the induced component is highly nontrivial. Without it, the displacement uncertainty, as well as the momentum uncertainty, will all decay with time such that the uncertainty principle can not be enforced. 
Fluctuation and dissipative backaction are two interlinked factors in consequence of interaction between the system and its environment. It manifests through a fluctuation-dissipation relation~\cite{HHL18} which guarantees that the uncertainty principle is satisfied.

\begin{figure}
	\centering
    \scalebox{0.5}{\includegraphics{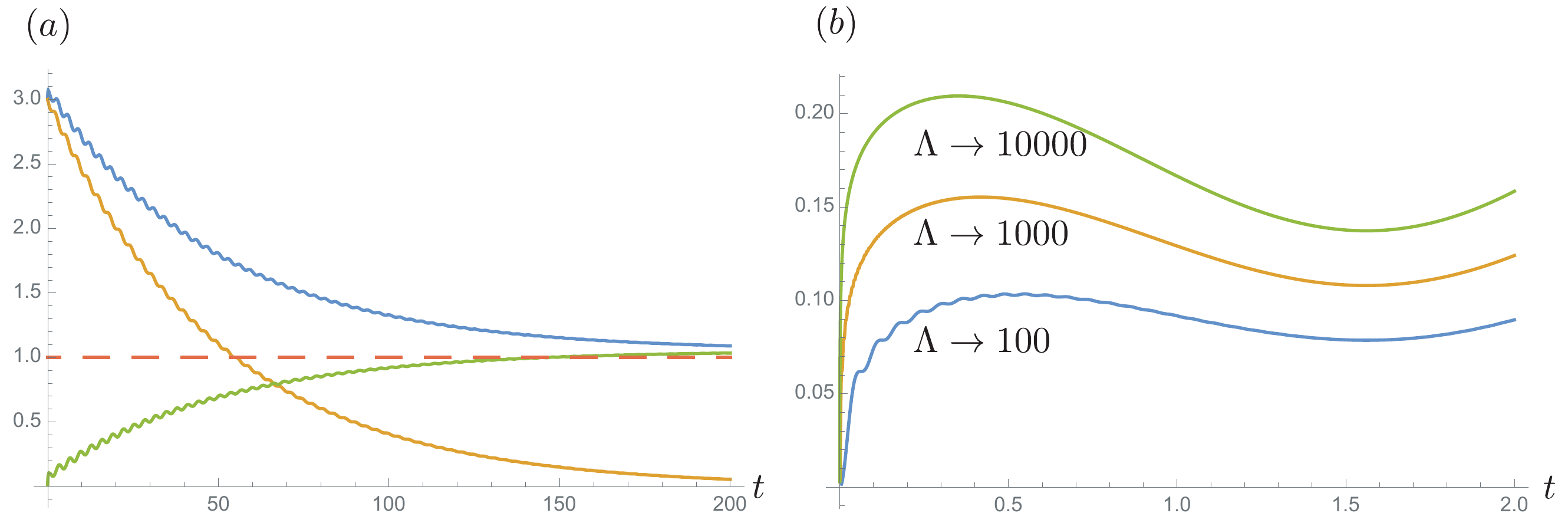}}
    \caption{In (a), we plot the time evolution of $\langle\hat{P}^{2}(t)\rangle$, normalized by {the corresponding value  of a free oscillator in the ground state}, for $\Lambda=100$, $\omega=1$, $\gamma=0.01$. In (b), we highlight the influence of the cutoff scale on the induced component, which is particularly visible at early time when a  predicted jolt shows up (a consequence of the assumption of a factorized initial state \cite{HPZ}). The initial jolt is steep when the cutoff scale is large.}\label{Fi:p2}
\end{figure}

We observe that for the induced component, the involvement of dissipative back-action at late times is crucial. The displacement uncertainty may grow linearly without bound if dissipation were not accounted for to balance the driving quantum noise of the environment field. This is another special feature of strong coupling predicted by using the full-fledged influence functional approach, amiss in results from time-dependent perturbation theory.

The cutoff scale barely has any effect on the displacement uncertainty but plays a more important role in the momentum uncertainty at strong coupling, as has been demonstrated in the analytical expression of the late-time value of the momentum uncertainty. If a larger cutoff scale is chosen in the theory, then typically it will cause a smooth but sharper rise in the value of the induced component of momentum uncertainty within the time scale of the order $\Lambda^{-1}$ right after the atom-field interaction is turned on. The is the so-called jolt phenomenon that typically occurs in the quantity that has cutoff dependence. This is also related to the $\gamma\ln\Lambda$ term in the late-time value of $\lambda$, so it will be more notable for strong atom-field interaction. Another effect of the cutoff scale will be manifested in the high-frequency wiggling, but its amplitude is roughly inversely proportional to $\Lambda$ and decays with time, so it is a minor effect. In either displacement or momentum uncertainty, the time scale $\omega^{-1}$ is less interesting unless it becomes comparable to the relaxation time scale $\gamma^{-1}$, as shown in Fig.~\ref{Fi:x2}-(b), where the linear growth of the induced component of $\langle\hat{Q}^{2}(t)\rangle$ at early times is harder to see due to the oscillatory feature of the period $\omega^{-1}$. In contrast, in Fig.~\ref{Fi:x2}-(a), since $\gamma^{-1}\gg\omega^{-1}$, the same oscillatory feature appears more like ripples on different curves.

\begin{figure}
	\centering
    \scalebox{0.6}{\includegraphics{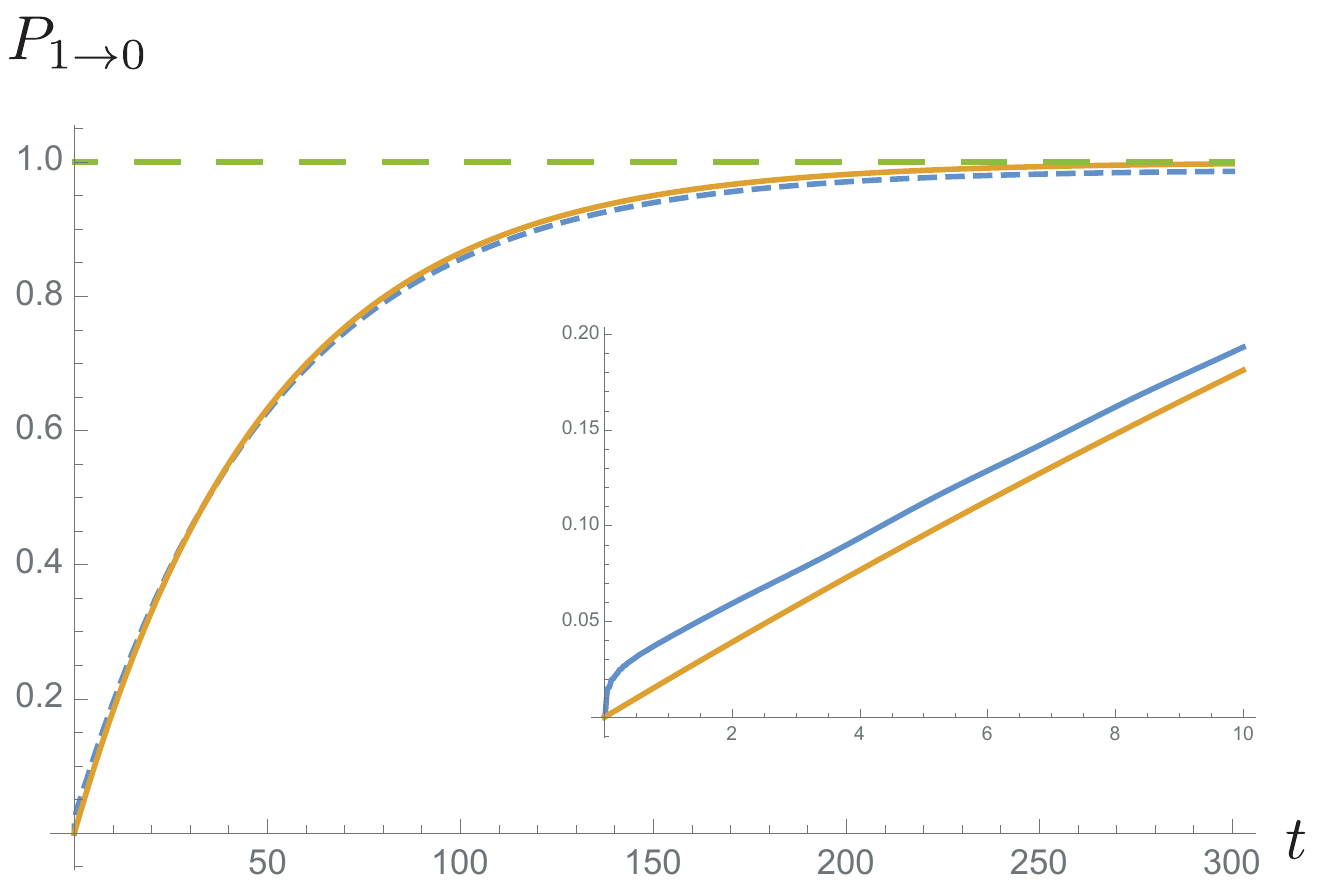}}
    \caption{The transition probability of spontaneous emission in the \textit{weak coupling} limit. The blue short-dashed curve is the result predicted by the influence functional method used here while the orange solid curve is the one given by using the time-dependent perturbation theory with Einstein's $A B$ coefficients. Both theories give comparable predictions by and large. Two subtle differences are shown in the regimes $t\gg\gamma^{-1}$ and $t\lesssim\Lambda^{-1}$. The inset highlights the region $t<\Lambda^{-1}$ with the prediction of the perturbation theory drawn in orange. The parameters are chosen to be $\Lambda=100$, $\omega=1$, $\gamma=0.01$.}\label{Fi:p10}
\end{figure}

Now we turn to the transient behavior of the transition probability of  spontaneous emission when the atom is initially in its first excited state. We assume that the ambient scalar field is in its vacuum state and the atom resides in an unbounded Minkowski spacetime. For weak atom-field coupling $\gamma\ll\omega$, among various length scales mentioned earlier, $\gamma^{-1}$ and $\Lambda^{-1}$ are particularly important. The former dictates the relaxation time, the amount of time it takes for the atom to reach equilibrium with the ambient scalar field, and the latter governs the time scale at which the correlation between the atom and the field is mixed and scrambled. Right after the interaction is switched on, the transition probability $P_{1\to0}$ grows like
\begin{equation}
	P_{1\to0}(t)\simeq\gamma t+\Bigl(\gamma^{2}+\frac{\gamma\Lambda^{2}}{2\pi\omega}\Bigr)\,t^{2}+\cdots\,,
\end{equation}
within the time   $0\leq t\leq\Lambda^{-1}$. Then it gradually turns to a gentle linear growth with the slope $2\gamma$, 
\begin{equation}\label{E:irtnsd}
	P_{1\to0}(t)\simeq2\gamma t+\cdots\,,
\end{equation}
when $\Lambda^{-1}\ll t\ll\gamma^{-1}$. Eventually it saturates to a time independent value \eqref{E:tjkwndsfs}, which is very close to unity in the limit of weak atom-field coupling. The linear growth \eqref{E:irtnsd} of the transition probability has been well understood by time-dependent  perturbation theory, in which the atom-field coupling is taken to be very weak and the atomic states are assumed unaltered, i.e., the back-action from the field is absent. In the infinite time limit, the Fermi Golden rule can be applied to find the transition probability. Unfortunately this approach predicts an infinite probability in this limit. However, if we assume a large yet finite time, then dividing the probability by this evolution time yields a finite transition probability rate $2\gamma$, as given in \eqref{E:irtnsd}.

The linear growth at short times and divergence at large times of the transition probability of spontaneous emission predicted by  time-dependent perturbation theory have a strong resemblance to the behavior of the position dispersion of the pollen in the conventional Brownian motion. This results from the fact that only the fluctuations from the environment is considered, but not the complementary effect of dissipation required by self-consistency. Here the same observation applies to the transition probability, since modification of atomic states due to the interaction with the ambient scalar field has not been properly taken into account. Thus the behavior of the transition probability of spontaneous emission predicted by time-perturbation theory will not show saturation to a constant value. This shortcoming, however, can be remedied by the phenomenological Einstein's $AB$ coefficients, as has been briefly discussed in \eqref{E:gkhshfd}. There if we identify the coefficient $A$ by $2\gamma$, then from \eqref{E:gvjhyerd} we see that the population $n_{g}(t)$ of the ground state, or equivalently the transition probability $P_{1\to0}$ to the ground state, will relax with time at a rate $2\gamma$, i.e.,
\begin{equation}
	n_{g}(t)=n\bigl(1-e^{-2\gamma t}\bigr)\,.
\end{equation}
In Fig.~\ref{Fi:p10}, we show the results of the transition probability of spontaneous emission by our approach in contrast to the combination of time-dependent perturbation theory and Einstein's $A B$ coefficients. In the weak coupling limit, they match very well, and the saturated values are very close to one. However, there are two major differences. In TDPT theory, the saturated value is always unity, while our approach indicates that the value is close to but not exactly equal to one; the deviation from unity is of the order $\gamma$. The second disparity resides  in the regime $t<\Lambda^{-1}$, where the probability curve rises very rapidly. This is associated with the jolt phenomenon, discussed earlier in the context of momentum dispersion. There are some other minor differences between results obtained from these two theories. When we zoom in to inspect the curve over the scale $\omega^{-1}$, we will see small oscillations of the same scale along the curve. If we examine even more closely, we will see higher-frequency wiggles of the period of order $\Lambda^{-1}$. Nonetheless, since these fast oscillations have very small amplitudes and decay with time, they are usually less important in the limit of weak atom-field interaction.

\begin{figure}
	\centering
    \scalebox{0.6}{\includegraphics{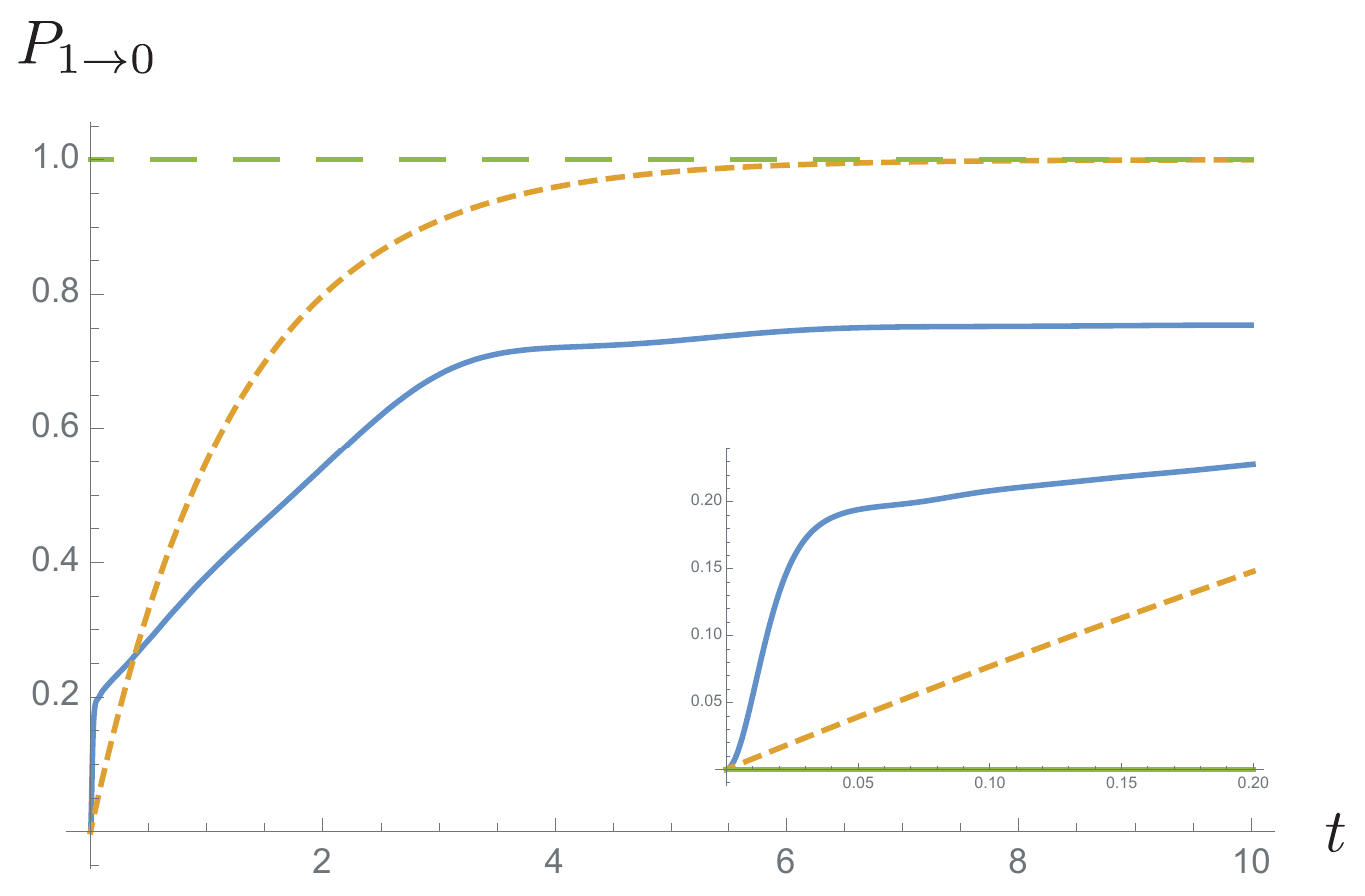}}
    \caption{The transition probability of spontaneous emission in the \textit{strong coupling} limit. The blue solid curve is the result predicted by the influence functional method used here while the orange short-dash curve is the one given by time-dependent perturbation theory. The  qualitative  difference is very pronounced. Linear growth sustains for the time $t<\omega^{-1}$ instead of $\gamma^{-1}$. The inset shows the region $t<20\Lambda^{-1}$. The parameters are chosen such that $\Lambda=100$, $\omega=1$, $\gamma=0.4$.}\label{Fi:p10str}
\end{figure}

The situation is drastically different in the strong coupling regime where $\gamma^{-1}$ and $\omega^{-1}$ are of the same order. Dynamics of two different time scales $\omega^{-1}$, $\gamma^{-1}$ have comparable contributions, so it is harder to see the linear growth of the transition probability as in the weak coupling case. In the event the damping constant $\gamma$ is at most a couple times smaller than $\omega$, we can still observe linear growth within the evolution time $\Lambda^{-1}<t<\omega^{-1}$. As shown in Fig.~\ref{Fi:p10str}, time-dependent perturbation theory fails miserably in the strong coupling regime. The linear growth of the transition probability in the transient regime does not have a rate $2\gamma$ as in the weak coupling case. Instead, it increases approximately at the rate
\begin{equation}
	\frac{2\gamma}{\Bigl[1+\dfrac{4\gamma}{\pi\omega}\Bigl(\ln\frac{\Lambda}{\omega}-1\Bigr)\Bigr]^{\frac{3}{2}}}<2\gamma\,,
\end{equation}
for $\Lambda^{-1}<t<\omega^{1}<\gamma^{-1}$. The rapid rise in the very beginning is again related to the jolt phenomenon, dependent on the cutoff scale in the transition probability.

As has been discussed earlier, when the strength of the atom-field interaction is increased, the behavior of the transition probability of spontaneous emission gradually deviates from the predictions based on time-dependent perturbation theory. In particular, the saturated value can be far below unity. This means that there is always a finite probability that the atom does not fall back to the ground state. We have to be careful in what  transition probability means. In the conventional (weak coupling) sense it is the probability from one energy level of a free atom to anther, not the probability of a free atom to the final state of the interacting atom. A projective measurement in the end is necessary. Let us explain.

In dealing with an atom in a quantized radiation field, we often write the states of the total system as the product of the eigenstates of the respective Hamiltonians of the {free} atom and the radiation field. The mutual interaction then introduces level shift and finite lifetime to the energy levels of the atom in the weak coupling limit, so that the spectral density has a Lorentz-function like distribution about the ``location'' of each level. The width of the distribution is related to the lifetime of the level, and is inversely proportional to the coupling strength $\gamma$. Thus the product state is conventionally seen as a good approximated description of the total system. However, strictly speaking, the product states cannot be the eigenstates of the Hamiltonian of the total system, because by definition the corresponding eigenstates should be stable, and thus have infinitely long life time. As we increase the atom-field interaction we expect that atomic levels will show a wider dispersion or a shorter lifetime to such an extent that the spectral density does not have any well-located peaks which define  the energy levels of the atom. The product state is not a good approximation of the eigenstate of the Hamiltonian of the total system. From the previous discussions, when we extend the evolution of the atomic system from the transient to the relaxation stages, we see the atom-field interaction modifies the atom's density matrix non-trivially. Thus the true eigenstate of the total system must include a superposition of various product states of the atom and the radiation to account for all sorts of effects due to strong interaction between them, and then the reduced description of the atom will inevitably become  mixed. If we initially prepare the atom in its first excited state before the atom-field  interaction is turned, then following the previous discussion when we repeatedly perform the projective measurements in terms of free atomic states, we will discover that the atom can stay in any of them with a certain probability; not necessarily destined to fall to its original ground state. 

\section{Summary}

{In this paper we highlight several features of atomic transitions in an atom strongly coupled to a quantum field in free space which are absent in traditional treatments under weak atom-field coupling.  We consider an atom whose internal degrees of freedom are modeled by a harmonic oscillator  which are  strongly coupled to a massless scalar quantum field. If their individual initial states are Gaussian, the dynamics of the entire system can be exactly solved, regardless of the coupling strength between the atom and field. In particular, we use the influence functional or the `in-in' formalism to calculate the reduced density matrix of the harmonic atom obtained by integrating over the  quantum field. It carries all the dynamical information of the harmonic atom under the influence of a quantum field. The features due to strong atom-field coupling include, but are not limited to,
	\begin{enumerate}
		\item Cutoff dependence: The contribution associated with the cutoff scale in the model may not be ignored. Since it takes the form as the product of the damping constant times the logarithm of the cutoff scale, what is seen as negligibly small in the weak coupling limit can become significant at strong coupling. Thus this contribution needs to be consistently taken into account.
		\item Non-Gibbsian density matrix: The density matrix of a strongly coupled  harmonic atom  does not have the Gibbsian form. Thus when it is placed in the ambient thermal field, its energy levels will not follow the canonical distribution. This has interesting new thermodynamical implications.
		\item Spectral density: Strong atom-field interaction will introduce very wide dispersion to the energy levels such that the originally discrete energy spectrum becomes continuous or gapless. This implies that the (reduced) relaxed, lowest-energy state of the atom can have energy fluctuations because this final state is not the energy eigenstate of the Hamiltonian of the free atom. 
		\item Ground-state spontaneous excitation: At strong atom-field coupling, even if both the atom and the field are initially in their ground state, there is a finite probability for the atom to transit to the excited states. This interesting process is not restricted to the ground state, but is most clearly seen for this case. It can be understood theoretically from the observation that the product state of the atom and the field is not an energy eigenstate of the entire system due to non-negligible contribution of the interaction term. Thus the product state of the ground state of each individual system is not a genuine ground state of the entire system. It gives a nonzero probability for the atom to be spontaneously excited to higher levels. This probability, roughly speaking, is proportional to the damping constant to the power which depends on the number of the levels the atom need to transit to. In the weak-coupling limit most of these transitions become forbidden, consistent with the conventional perturbative treatment.
		\item Non-exponential relaxation of the transition probability: At strong coupling, the time scale associated with the inverse of the damping constant $\gamma$ is not the dominant one as in the weak-coupling case; it can be comparable with the time scale related to the inverse of the energy level difference. This implies that the time evolution of the transition probability curve will not exponentially approach  to its final value; additional structures will emerge due to the competitive behaviors between time scales associated with the parameters in the model. 
	\end{enumerate}

\noindent {\bf Acknowledgments} Part of this work was done when JTH visited the Maryland Center for Fundamental Physics and BLH visited the Institute of Physics at the Academia Sinica, Taiwan, ROC.

\appendix


\section{Differences between TDPT and IF results for the transition probability}

The model used in this paper is exactly solvable by the influence functional (IF) formalism. It thus offers a nice platform to check on the deficiencies of any method which makes approximations, such as the Markovian approximation, the rotating wave approximation, Born-Markov approximation, Lindblad-Redfield equations, Fermi Golden rule etc. Here we compare the results for the transition probabilities obtained from the commonly used time-dependent perturbation theory (TDPT)  against the exact solutions we obtained via the IF method, point out where it erred and the parameter regimes where the deficiencies lie.

Consider a detector, whose internal degree of freedom $Q(t)$ is modeled by a harmonic oscillator and is coupled to a massless quantum scalar field $\phi(x)$. Assume that initially the detector is in an excited state $\lvert E_{m}\rangle_{\textsc{a}}$, with energy $E_{m}$ of a free oscillator, while the scalar field is in a vacuum state $\lvert0\rangle_{\textsc{f}}$. We are interested in the detector's transition probability upon interacting with the quantum field. Since this model is exactly solvable when the detector-field interaction is bilinear, no matter how strong the the coupling strength is, we will compare the disparities, predicted by two different approaches: 1) time-dependent perturbation theory and 2) Feynmann-Vernon influence functional formalism.

Suppose that the total Hamiltonian operator of this system can be formally decomposed into $\hat{H}^{(\textsc{s})}=\hat{H}_0^{(\textsc{s})}+\lambda\,\hat{V}^{(\textsc{s})}$, where the superscript s indicates that we are working in the Schr\"odinger picture. The subscript 0 indicates that the quantity corresponds to the free case, e.g,  $\hat{H}_0^{(\textsc{s})}$ contains the free Hamiltonians of the detector's internal degree of freedom and the scalar field. The interaction term $\hat{V}^{(\textsc{s})}$ is attached with a bookkeeping tag $\lambda$, which can also serve as the coupling strength when needed.  In the interaction picture, the evolution of the state of the total system $\lvert\Psi^{(\textsc{i})}(t)\rangle$ from time $t_i$ to $t_f$ is governed by
\begin{align}\label{E:uthddf}
	\lvert\Psi^{(\textsc{i})}(t_f)\rangle&=\hat{U}^{(\textsc{i})}(t_f,t_i)\,\lvert\Psi^{(\textsc{i})}(t_i)\rangle\,,&\hat{U}^{(\textsc{i})}(t_f,t_i)&=\operatorname{T}_{+}\exp\biggl\{-i\,\lambda\int_{t_i}^{t_f}\!dt''\;\hat{V}^{(\textsc{i})}(t'')\biggr\}\,,
\end{align}
where $\hat{U}^{(\textsc{i})}(t_f,t_i)$ is the unitary evolution operator in the interaction picture, denoted by the superscript $(\textsc{i})$. The operator $\operatorname{T}_{+}$ tells us that the expression afterwards will be time-ordered. In the weak coupling case, we may Taylor expand \eqref{E:uthddf} and arrive at
\begin{align}
	\lvert\Psi^{(\textsc{i})}(t_{f})\rangle=\biggl\{1-i\,\lambda\int^{t_{f}}_{t_{i}}\!dt'\;\hat{V}^{(\textsc{i})}(t')-\lambda^{2}\int^{t_{f}}_{t_{i}}\!dt'\!\int^{t'}_{t_{i}}\!dt''\;\hat{V}^{(\textsc{i})}(t')\hat{V}^{(\textsc{i})}(t'')+\mathcal{O}(\lambda^{3})\biggr\}\,\lvert\Psi^{(\textsc{i})}(t_{i})\rangle\,.
\end{align}
Now suppose the initial state $\lvert\Psi^{(\textsc{i})}(t_{i})\rangle$ is $\lvert E_m^{(\textsc{i})}\rangle_{\textsc{a}}\lvert0^{(\textsc{i})}\rangle_{\textsc{f}}$ and the final state $\lvert\Psi^{(\textsc{i})}(t_{f})\rangle$ is $\lvert E_n^{(\textsc{i})}\rangle_{\textsc{a}}\lvert\gamma^{(\textsc{i})}\rangle_{\textsc{f}}$, where $\lvert\gamma^{(\textsc{i})}\rangle_{\textsc{f}}$ represents some particle number state of the scalar field in the interaction picture. If we are interested only in the transition probability of the detector from $\lvert E_m^{(\textsc{i})}\rangle_{\textsc{a}}$ to $\lvert E_n^{(\textsc{i})}\rangle_{\textsc{a}}$, then such a probability will be given by
\begin{align}
	P_{m\to n}(t_{f})&=\sum_{\lvert\gamma^{\textsc{(i)}}\rangle_{\textsc{f}}}\,\Bigl|\;{}_{\textsc{a}}\langle E^{\textsc{(i)}}_{n}\rvert\,{}_{\textsc{f}}\langle\gamma^{\textsc{(i)}}\rvert\,\biggl[1-i\,\lambda\int\!d^{3}x'\int_{t_{i}}^{t_{f}}\!dt'\;\hat{V}^{\textsc{(i)}}(x',t')+\cdots\biggr]\,\rvert E^{\textsc{(i)}}_{m}\rangle_{\textsc{a}}\lvert0^{\textsc{(i)}}\rangle_{\textsc{f}}\;\Bigr|^{2}\notag\\
	&=\sum_{\lvert\gamma^{\textsc{(i)}}\rangle_{\textsc{f}}}\biggl\{{}_{\textsc{a}}\langle E^{\textsc{(i)}}_{m}\rvert\,{}_{\textsc{f}}\langle0^{\textsc{(i)}}\rvert\,\biggl[1+i\,\lambda\int\!d^{3}x''\int_{t_{i}}^{t_{f}}\!dt''\;\hat{V}^{\textsc{(i)}}{}^{\dagger}(x'',t'')+\cdots\biggr]\,\rvert E^{\textsc{(i)}}_{n}\rangle_{\textsc{a}}\lvert\gamma^{\textsc{(i)}}\rangle_{\textsc{f}}\biggr.\notag\\
	&\qquad\times\biggl.{}_{\textsc{a}}\langle E^{\textsc{(i)}}_{n}\rvert\,{}_{\textsc{f}}\langle\gamma^{\textsc{(i)}}\rvert\,\biggl[1-i\,\lambda\int\!d^{3}x'\int_{t_{i}}^{t_{f}}\!dt'\;\hat{V}^{\textsc{(i)}}(x',t')+\cdots\biggr]\,\rvert E^{\textsc{(i)}}_{m}\rangle_{\textsc{a}}\lvert0^{\textsc{(i)}}\rangle_{\textsc{f}}\biggr\}\,,\label{E:eirhd}
\end{align}
after we have summed over all of the intermediate particle number states of the field.

Let us assume that the interaction term $\hat{V}$ is given by the typical bilinear coupling $\hat{Q}(t)\hat{\phi}(t)$ in $1+3$-dimensional flat spacetime with $\hat{\phi}(t)\equiv\hat{\phi}(\mathbf{z},t)$, where $\mathbf{z}$ is the prescribed spatial location of the detector. If the initial state and the final state of the detector are orthogonal, i.e., ${}_{\textsc{a}}\langle E_{n}\vert E_{m}\rangle_{\textsc{a}}=0$, then we only need to examine the term of the order $\lambda^{2}$ in \eqref{E:eirhd}. Since we would like to compare this result with that of the influence functional formalism, we will rotate the states and the operators of the detector back to their counterparts in the Schr\"odinger picture, and \eqref{E:eirhd} becomes
\begin{align}
	P_{m\to n}^{(2)}(t_{f})&=\lambda^{2}\int_{t_{i}}^{t_{f}}\!dt'\!\int_{t_{i}}^{t_{f}}\!dt''\;{}_{\textsc{a}}\langle E^{\textsc{(s)}}_{0}\rvert\,\hat{U}_{Q',0}^{\dagger}(t'',t_{i})\,\hat{Q}'^{(\textsc{s})}(t'')\,U_{Q',0}^{\dagger}(t_{f},t'')\,\rvert E^{\textsc{(s)}}_{n}\rangle_{\textsc{a}}\label{E:bvhgsd}\\
	&\qquad\times{}_{\textsc{a}}\langle E^{\textsc{(s)}}_{n}\rvert\,U_{Q,0}^{\vphantom{\dagger}}(t_{f},t')\,\hat{Q}^{\textsc{(s)}}(t')\,U_{Q,0}^{\vphantom{\dagger}}(t',t_{i})\,\rvert E^{\textsc{(s)}}_{0}\rangle_{\textsc{a}}\times{}_{\textsc{f}}\langle0^{\textsc{(i)}}\rvert\,\hat{\phi}'^{\textsc{(i)}}{}^{\dagger}(t'')\hat{\phi}^{\textsc{(i)}}(t')\,\lvert0^{\textsc{(i)}}\rangle_{\textsc{f}}\,.\notag
\end{align}
The superscript $(2)$ on the transition probability $P_{m\to n}$ indicates that the contribution is of the order $\lambda^{2}$. The operator $\hat{U}_{Q,0}(t,t')$ is the unitary time evolution operator of the free detector, and is used to transform the detector between the Schr\"odinger and interaction pictures,
\begin{align}
    \lvert E_m^{(\textsc{i})}(t)\rangle_{\textsc{a}}&=\hat{U}_{Q,0}^{\dagger}(t,t_0)\,\lvert E_m^{(\textsc{s})}(t)\rangle_{\textsc{a}}\,,&\hat{O}_{Q}^{(\textsc{i})}(t)&=\hat{U}_{Q,0}^{\dagger}(t,t_0)\,\hat{O}_{Q}^{(\textsc{s})}(t)\,\hat{U}_{Q,0}^{\vphantom{\dagger}}(t,t_0)\,.
\end{align}
Here $\hat{O}_{Q}$ is the operator associated with the detector and we have assumed that both pictures coincide at time $t_0$. The term ${}_{\textsc{f}}\langle0^{\textsc{(i)}}\rvert\,\hat{\phi}'^{\textsc{(i)}}{}^{\dagger}(t'')\hat{\phi}^{\textsc{(i)}}(t')\,\lvert0^{\textsc{(i)}}\rangle_{\textsc{a}}$ is independent of the pictures and turns out to be the Schwinger function $-i\,G_{-+}^{(\phi)}(t'',t')$ of the free scalar field. Now let $\psi_{n}^{*}(Q_{f},t_{f})={}_{\textsc{a}}\langle E^{\textsc{(s)}}_{n}(t_{f})\vert Q_{f}\rangle$ and $\psi_{m}^{\vphantom{*}}(Q_{i},t_{i})=\langle Q_{i}\vert E^{\textsc{(s)}}_{m}(t_{i})\rangle_{\textsc{a}}$. The initial density matrix elements of the detector is then given by $\hat{\rho}_{Q}(Q^{\vphantom{'}}_{i},Q'_{i};t_{i})=\psi_{m}^{\vphantom{*}}(Q_{i},t_{i})\psi_{m}^{*}(Q'_{i},t_{i})$, and Eq.~\eqref{E:bvhgsd} becomes
\begin{align}\label{E:erbfhd}
	&\quad P_{m\to n}^{(2)}(t_{f})\\
	&=-i\,\lambda^{2}\int\!dQ^{\vphantom{'}}_{f}\,dQ'_{f}\;\psi_{n}^{*}(Q_{f},t_{f})\psi_{n}^{\vphantom{*}}(Q'_{f},t_{f})\int\!dQ_{i}\,dQ'_{i}\;\rho_{Q}(Q^{\vphantom{'}}_{i},Q'_{i};t_{i})\int_{t_{i}}^{t_{f}}\!dt'\!\int_{t_{i}}^{t_{f}}\!dt''\;G_{-+}^{(\phi)}(t'',t')\notag\\
	&\quad\times\langle Q_{f}\rvert\,\hat{U}_{Q,0}^{\vphantom{\dagger}}(t_{f},t')\,\hat{Q}^{\textsc{(s)}}(t')\,\hat{U}_{Q,0}^{\vphantom{\dagger}}(t',t_{i})\,\rvert Q_{i}\rangle\times\langle Q'_{i}\rvert\,\hat{U}_{Q',0}^{\dagger}(t'',t_{i})\,\hat{Q}'^{(\textsc{s})}(t'')\,\hat{U}_{Q',0}^{\dagger}(t_{f},t'')\,\lvert Q'_{f}\rangle\,.\notag
\end{align}
We will use this expression to compare with the corresponding result obtained by the influence functional formalism.

The transition probability from the state $\lvert\Psi_{i}\rangle$ and the state $\lvert\Psi_{f}\rangle$ can be expressed in terms of the density matrix,
\begin{equation}\label{E:gggdrre}
	P_{i\to f}=\operatorname{Tr}\Bigl\{\lvert\Psi_{f}\rangle\langle\Psi_{f}\rvert\,\hat{\rho}_{\Psi}(t_{f})\Bigr\},
\end{equation}
where the density matrix $\rho_{\Psi}$ at $t=t_{f}$ is given by
\begin{equation}
	\hat{\rho}_{\Psi}(t_{f})=\hat{U}(t_{f},t_{i})\,\lvert\Psi_{i}\rangle\langle\Psi_{i}\rvert\,{U}^{\dagger}(t_{f},t_{i})=\hat{U}(t_{f},t_{i})\,\hat{\rho}_{\psi}(t_{i})\,\hat{U}^{\dagger}(t_{f},t_{i})\,.
\end{equation}
Here $\hat{U}$ is the unitary time evolution operator that governs the dynamics of $\lvert\Psi\rangle$. For an  open quantum system composed of a detector and a massless scalar field, the reduced density matrix of the internal degree of freedom $Q$ of the detector is given by $\hat{\rho}_{Q}=\operatorname{Tr}_{\phi}\hat{\rho}_{Q\phi}$.  In a path integral representation its elements take the form
\begin{align}
	\rho_{Q}(Q^{\vphantom{'}}_{f},Q'_{f};t_{f})&=\operatorname{Tr}_{\phi}\Bigl\{\hat{U}(t_{f},t_{i})\,\hat{\rho}_{Q\phi}(t_{i})\,\hat{U}^{\dagger}(t_{f},t_{i})\Bigr\}\label{E:eoiruhfd}\\
	&=\int_{-\infty}^{\infty}\!dQ^{\vphantom{'}}_{i}\,dQ'_{i}\;\rho_{Q}(Q^{\vphantom{'}}_{i},Q'_{i};t_{i})\int_{Q_{i}}^{Q_{f}}\!\mathcal{D}Q_{+}\int_{Q'_{i}}^{Q'_{f}}\!\mathcal{D}Q_{-}\label{E:ebfjds}\\
	&\quad\times\exp\biggl\{i\,S_{Q}[Q_{+}]-i\,S_{Q}[Q_{-}]+\frac{i}{2}\int_{t_{i}}^{t_{f}}\!d^{4}x\,d^{4}x'\sum_{a,b=\pm}J_{a}(x)\,G_{ab}^{(\phi)}(x^{\mu},x'^{\mu})\,J_{b}(x')\biggr\}\,,\notag
\end{align}
where $x^{\mu}=(t,\mathbf{x})$, $J_{a}(x^{\mu})=\lambda\,Q_{a}(t)\,\delta^{(3)}(\mathbf{x}-\mathbf{z})$. Here $S_{Q}$ is the action of the free detector, and $G_{ab}^{(\phi)}$ is a collection of various Green's functions of the free scalar field~\cite{Greiner,CalHu},
\begin{align*}
	G_{++}^{(\phi)}(x^{\mu},x'^{\mu})&=G_{F}^{(\phi)}(x^{\mu},x'^{\mu})=i\,\langle\,\operatorname{T}_{+}\hat{\phi}(x^{\mu})\hat{\phi}(x')\,\rangle\,,\\
	G_{--}^{(\phi)}(x^{\mu},x'^{\mu})&=G_{D}^{(\phi)}(x^{\mu},x'^{\mu})=i\,\langle\,\operatorname{T}_{-}\hat{\phi}(x^{\mu})\hat{\phi}(x'^{\mu})\,\rangle\,,\\
	G_{-+}^{(\phi)}(x^{\mu},x'^{\mu})&=G_{>}^{(\phi)}(x^{\mu},x'^{\mu})=i\,\langle\,\hat{\phi}(x^{\mu})\hat{\phi}(x'^{\mu})\,\rangle\,,\\
	G_{+-}^{(\phi)}(x^{\mu},x'^{\mu})&=G_{<}^{(\phi)}(x^{\mu},x'^{\mu})=i\,\langle\,\hat{\phi}(x'^{\mu})\hat{\phi}(x^{\mu})\,\rangle\,,
\end{align*}
where $\operatorname{T}_{-}$ denotes anti-time-ordering. In particular $G_{F}^{(\phi)}(x^{\mu},x'^{\mu})$ and $G_{D}^{(\phi)}(x^{\mu},x'^{\mu})$ are the Feynman and Dyson propagators.

To establish a direct comparison with the result \eqref{E:erbfhd} of time-dependent perturbation theory, we can Taylor-expand the expression in the influence functional in \eqref{E:ebfjds} that contains the source $J_a$ by assuming a weak detector-field interaction. We find that the reduced density at time $t_{f}$ is hence given by
\begin{align}
	\rho_{Q}(Q^{\vphantom{'}}_{f},Q'_{f};t_{f})&=\int_{-\infty}^{\infty}\!dQ^{\vphantom{'}}_{i}\,dQ'_{i}\;\rho_{Q}(Q^{\vphantom{'}}_{i},Q'_{i};t_{i})\int_{Q_{i}}^{Q_{f}}\!\mathcal{D}Q_{+}\int_{Q'_{i}}^{Q'_{f}}\!\mathcal{D}Q_{-}\,\exp\biggl\{i\,S_{Q}[Q_{+}]-i\,S_{Q}[Q_{-}]\biggr\}\notag\\
	&\qquad\qquad\quad\times\biggl\{1+i\,\frac{\lambda^{2}}{2}\int_{t_{i}}^{t_{f}}\!dt'\,dt''\;Q_{+}(t')\,G_{++}^{(\phi)}(t',t'')\,Q_{+}(t'')\biggr.\notag\\
	&\qquad\qquad\qquad\qquad+i\,\frac{\lambda^{2}}{2}\int_{t_{i}}^{t_{f}}\!dt'\,dt''\;Q_{-}(t')\,G_{--}^{(\phi)}(t',t'')\,Q_{-}(t'')\notag\\
	&\qquad\qquad\qquad\qquad-\biggl.i\,\lambda^{2}\int_{t_{i}}^{t_{f}}\!dt'\,dt''\;Q_{-}(t')\,G_{-+}^{(\phi)}(t',t'')\,Q_{+}(t'')+\cdots\biggr\}\,.\label{E:rtnfjgnd}
\end{align}
Since we assume the final state of the detector is orthogonal to the initial state, we observe
\begin{align}
	\int\!dQ_{f}\,dQ'_{f}\;\psi_{n}^{*}(Q_{f},t_{f})\psi_{n}^{\vphantom{*}}(Q'_{f},t_{f})&\int_{-\infty}^{\infty}\!dQ^{\vphantom{'}}_{i}\,dQ'_{i}\;\rho_{Q}(Q^{\vphantom{'}}_{i},Q'_{i};t_{i})\label{E:gbwuers1}\\
	\times&\int_{Q_{i}}^{Q_{f}}\!\mathcal{D}Q_{+}\int_{Q'_{i}}^{Q'_{f}}\!\mathcal{D}Q_{-}\,\exp\biggl\{i\,S_{Q}[Q_{+}]-i\,S_{Q}[Q_{-}]\biggr\}=0\,,\notag
\intertext{and}
	\int\!dQ_{f}\,dQ'_{f}\;\psi_{n}^{*}(Q_{f},t_{f})\psi_{n}^{\vphantom{*}}(Q'_{f},t_{f})&\int_{-\infty}^{\infty}\!dQ^{\vphantom{'}}_{i}\,dQ'_{i}\;\rho_{Q}(Q^{\vphantom{'}}_{i},Q'_{i};t_{i})\label{E:gbwuers2}\\
	\times\int_{Q_{i}}^{Q_{f}}\!\mathcal{D}Q_{+}\int_{Q'_{i}}^{Q'_{f}}\!\mathcal{D}Q_{-}\,\exp\biggl\{&i\,S_{Q}[Q_{+}]-i\,S_{Q}[Q_{-}]\biggr\}\int_{t_{i}}^{t_{f}}\!dt'\,dt''\;Q(t')\,G_{++}^{(\phi)}(t',t'')\,Q(t'')=0\,,\notag
\end{align}
This is because in both \eqref{E:gbwuers1} and \eqref{E:gbwuers2} for the branch $(-)$ that goes backward in time, the initial state $\lvert E_m\rangle_{\textsc{a}}$ undergoes a free evolution, so it will be orthogonal to the final state $\lvert E_n\rangle_{\textsc{a}}$ if initially both are already orthogonal, i.e., ${}_{\textsc{a}}\langle E_n(t_i)\vert E_m(t_i)\rangle_{\textsc{a}}=0$. The same argument  applies to the term involving $G_{--}^{(\phi)}$.

The transition probability is then given by
\begin{align}\label{E:eerfd}
	P_{m\to n}^{(2)}(t_{f})&=\int\!dQ_{f}\,dQ'_{f}\;\psi_{n}^{*}(Q_{f},t_{f})\psi_{n}^{\vphantom{*}}(Q'_{f},t_{f})\,\rho_{Q}(Q^{\vphantom{'}}_{f},Q'_{f};t_{f})\notag\\
	&=-i\,\lambda^{2}\int\!dQ_{f}\,dQ'_{f}\;\psi_{n}^{*}(Q_{f},t_{f})\psi_{n}^{\vphantom{*}}(Q'_{f},t_{f})\int_{-\infty}^{\infty}\!dQ^{\vphantom{'}}_{i}\,dQ'_{i}\;\rho_{Q}(Q^{\vphantom{'}}_{i},Q'_{i};t_{i})\\
	\times&\int_{Q_{i}}^{Q_{f}}\!\mathcal{D}Q_{+}\int_{Q'_{i}}^{Q'_{f}}\!\mathcal{D}Q_{-}\,\exp\biggl\{i\,S_{Q}[Q_{+}]-i\,S_{Q}[Q_{-}]\biggr\}\int_{t_{i}}^{t_{f}}\!dt'\,dt''\;Q'(t')\,G_{-+}^{(\phi)}(t',t'')\,Q(t'')\,,\notag
\end{align}
which is exactly \eqref{E:erbfhd}. Thus we have explicitly shown how the same perturbative results of the transition probability can be obtained both by the time-dependent perturbation theory and the influence functional method. The interesting part is what has the TDPT missed out.

The congenital disadvantage of TDPT lies in the assumption that the interaction between two subsystems must be sufficiently weak, so many new features of the strong coupling regime described in this paper are absent by default. Even in the weak coupling regime, it is quite well-known that the result given by the perturbation theory does not perform well at late times, even when higher-order corrections are considered. Secular terms will emerge at late time, and many algorithms such as the multi-scale and dynamical renormalization group approaches~\cite{CG94,CG96,OK10} are developed in order to recover the asymptotic behavior. This secular behavior has been seen, for example, in the theoretical modeling of the Brownian motion, where the velocity of the system (say, pollen particle) will grow indefinitely if the backactions from the environment (say, water molecules) are not fully taken into consideration. A better-known example in this context is the calculation of the atomic transition probability due to coupling of the atom with the electromagnetic field. The first-order time-dependent perturbation theory plus the Fermi Golden rule will give an infinite transition probability at late time; only the transition rate is well defined. It is fair to say that the time-dependent perturbation theory can give a decent description only  at early time for the transient dynamics.

As has been discussed earlier, when the internal degree of freedom, modeled by a quantum harmonic oscillator, of the detector is coupled to a quantum scalar field, part of the self-energy of the scalar field will contribute to the renormalization of the natural frequency. As is known in renormalization theory the physical values of the renormalized quantities run with energy and thus depend on the energy scale involved in the experiment and the actual values will be determined by the experimental measurements. On top of that, quantum fluctuations of the field act like a stochastic forcing term, a quantum noise, which adds to driving the motion of the oscillator, engendering quantum radiation~\cite{JohHu00,LinHu06}. The self-force from the reaction of this radiation will provide  a damping mechanism to the oscillator's motion~\cite{GalHu05,HHL18}. Thus in general the detector's internal degree of freedom no longer experiences a free oscillatory motion, instead it acts like a driven damped oscillator. Its dynamics will relax in time to an equilibrium state, ultimately governed by the noise force. This equilibration is possible on account of the balance between the fluctuations and the dissipation effects both on account of its coupling to the quantum field. In  time-dependent perturbation theory, since the interaction is treated as a perturbation, the state of the detector will evolve as a free oscillator, as seen for example in \eqref{E:gbwuers1} and \eqref{E:gbwuers2}, where the evolution is governed by $S_{Q}$, the action of the free oscillator. Theoretically, the frequency appearing in $S_{Q}$ is the bare frequency, but in the practice of time-dependent perturbation theory, it is implicitly assumed to be the physical frequency.
This may not pose any serious issue because renormalization is expected to happen in an extremely short time scale right after the interaction is turned on. Thus, practically, it will be the physical parameter that will be engaged in the ensuing dynamics. A more serious concern  comes from the other contributions from the coupling with the field: the noise force and the self-force. These two are self-consistently accounted for in the influence functional, but are lacking in the time-dependent perturbation theory. Thus we do not see in the perturbation theory any relaxation dynamics for the detector's internal degree of freedom. In turth, the relaxed final state of the detector is governed by the quantum fluctuations of the field, which is independent of and could be quite different from the initial state. This observation brings forth an incompatibility issue inherent in the perturbative calculation. That is, the following two procedures of finding the final steady-state behavior of the detector are not equivalent: a) to carry out the full, exact calculations first and then perform the perturbative expansion, versus b) to perform the perturbative expansion first and then carry out the subsequent calculations. The latter is the strategy usually taken in time-dependent perturbation theory. Therefore, the late-time results obtained by these two approaches need not be the same even in the weak coupling limit. This disparity is more severe when the detector-field coupling is not weak, as expounded in this paper.


\end{document}